\documentclass[sn-mathphys,Numbered]{sn-jnl}
\usepackage{graphicx}%
\usepackage{caption}
\usepackage{subcaption}
\usepackage{multirow}%
\usepackage{amsmath,amssymb,amsfonts}%
\usepackage{amsthm}%
\usepackage{mathrsfs}%
\usepackage{hyperref}
\usepackage[title]{appendix}%
\usepackage{xcolor}%
\usepackage{textcomp}%
\usepackage{manyfoot}%
\usepackage{booktabs}%
\usepackage{algorithm}%
\usepackage{algorithmicx}%
\usepackage{algpseudocode}%
\usepackage{listings}%
\usepackage{algorithm,algpseudocode}
\theoremstyle{thmstyleone}%
\theoremstyle{thmstyletwo}%
\theoremstyle{thmstylethree}%
\usepackage{xr-hyper}
\begin{document}


\title[Article Title]{A conditional latent autoregressive recurrent model for generation and forecasting beam dynamics in particle accelerators}

\author*[1]{\fnm{Mahindra} \sur{Rautela}}\email{mrautela@lanl.gov}
\author[1,2]{\fnm{Alan} \sur{Williams}}\email{awilliams@lanl.gov}
\author*[1]{\fnm{Alexander} \sur{Scheinker}}\email{ascheink@lanl.gov}

\affil[1]{\orgdiv{Applied Electrodynamics Group (AOT-AE)}, \orgname{Los Alamos National Laboratory}, \orgaddress{ \city{Los Alamos, NM}, \postcode{87547}, \country{USA}}}
\affil[2]{\orgdiv{Department of Mechanical and Aerospace Engineering}, \orgname{University of California}, \orgaddress{ \city{San Diego, CA}, \postcode{92093}, \country{USA}}}

\abstract{Particle accelerators are complex systems that focus, guide, and accelerate intense charged particle beams to high energy. Beam diagnostics present a challenging problem due to limited non-destructive measurements, computationally demanding simulations, and inherent uncertainties in the system. We propose a two-step unsupervised deep learning framework named as Conditional Latent Autoregressive Recurrent Model (CLARM) for learning the spatiotemporal dynamics of charged particles in accelerators. CLARM consists of a Conditional Variational Autoencoder (CVAE) transforming six-dimensional phase space into a lower-dimensional latent distribution and a Long Short-Term Memory (LSTM) network capturing temporal dynamics in an autoregressive manner. The CLARM can generate projections at various accelerator modules by sampling and decoding the latent space representation. The model also forecasts future states (downstream locations) of charged particles from past states (upstream locations). The results demonstrate that the generative and forecasting ability of the proposed approach is promising when tested against a variety of evaluation metrics.}


\maketitle

\section{Introduction}\label{sec1}
Many problems in physics have spatiotemporal dynamics, with dependent variables varying as a function of both spatial and temporal coordinates.  In this work, we consider a state space density $\rho(\mathbf{x},t)$ of a high dimensional dynamic system ($\mathbf{x}\in\mathbb{R}^n$) evolving according to dynamics
\begin{equation}
    \frac{\partial \rho}{\partial t} = F(\rho,\mathbf{x},\mathbf{P}(\mathbf{x},t),t), \label{spatiotemporal}
\end{equation}
where the system's state is influenced by external time-varying parameters $\mathbf{P}(\mathbf{x},t)$. One example of such a system is the beam density $\rho$ of a charged particle beam in an accelerator, where $\mathbf{x}$ represents particle positions and velocities and $\mathbf{P}(\mathbf{x},t)$ represents various time-varying electromagnetic devices that influence the beam's dynamic evolution \cite{scheinker2018demonstration}. Another example is the dynamics of weather, where $\mathbf{x}$ represents location and $\mathbf{P}(\mathbf{x},t)$ represents local environmental influences on the weather dynamics \cite{amato2020novel}. The study of such problems can be computationally challenging when complex temporal evolution is coupled with detailed spatial features resulting in a requirement of a fine mesh of both space and time \cite{zhou2022neural}.

With advancements in parallel computations using modern graphical processing units (GPUs), machine learning has opened new frontiers for solving such complex problems in physics. Some applications include fluid dynamics \cite{vinuesa2022enhancing}, acoustics \cite{rautela2022inverse}, astrophysics \cite{huerta2021accelerated}, nuclear physics \cite{boehnlein2022colloquium}, plasma physics \cite{gonoskov2019employing}, and biophysics \cite{alquraishi2021differentiable}. The majority of currently deployed deep learning (DL) techniques fall into either spatial or sequential learning, with limited emphasis on spatiotemporal dynamics \cite{reichstein2019deep}. While DL techniques designed to model spatiotemporal dynamical phenomena do exist, they are relatively scarce. Three-dimensional Convolutional Neural Networks (3DCNN) with hard physics constraints have been developed for solving the spatiotemporal Maxwell's equations for electrodynamics \cite{scheinker2023physics}. 3DCNNs have also been employed to capture spatiotemporal dynamics of the compressible Navier–Stokes equations by extracting spatial and temporal features simultaneously through a shared 3D kernel \cite{wandel2021teaching}. Convolutional Long Short-Term Memory (ConvLSTM) represents another widely used approach for handling spatiotemporal data. In this method, a CNN extracts spatial features, while an LSTM network focuses on temporal predictions \cite{shi2015convolutional}. The temporal learning is performed on a higher-dimensional feature space extracted by the CNN. Such a model has been used for short time precipitation forecasting (nowcasting) problems. Deep Convolutional Generative Adversarial Networks (DCGAN) have been used to leverage deep convolutional networks in an adversarial framework to learn flow dynamics \cite{cheng2020data}. DCGAN is implemented to accurately predict the flow field in test cases of tsunamis, demonstrating comparable results to the original high-fidelity model. Graph Neural Networks (GNN) are used for learning spatiotemporal data and are utilized for the prediction of fluid flow physics problems \cite{kipf2016semi}. GNN have also been applied for surrogate modeling to predict the laminar flow around two-dimensional bodies \cite{chen2021graph}.

\begin{figure}
    \centering
    \includegraphics[width=1.0\textwidth]{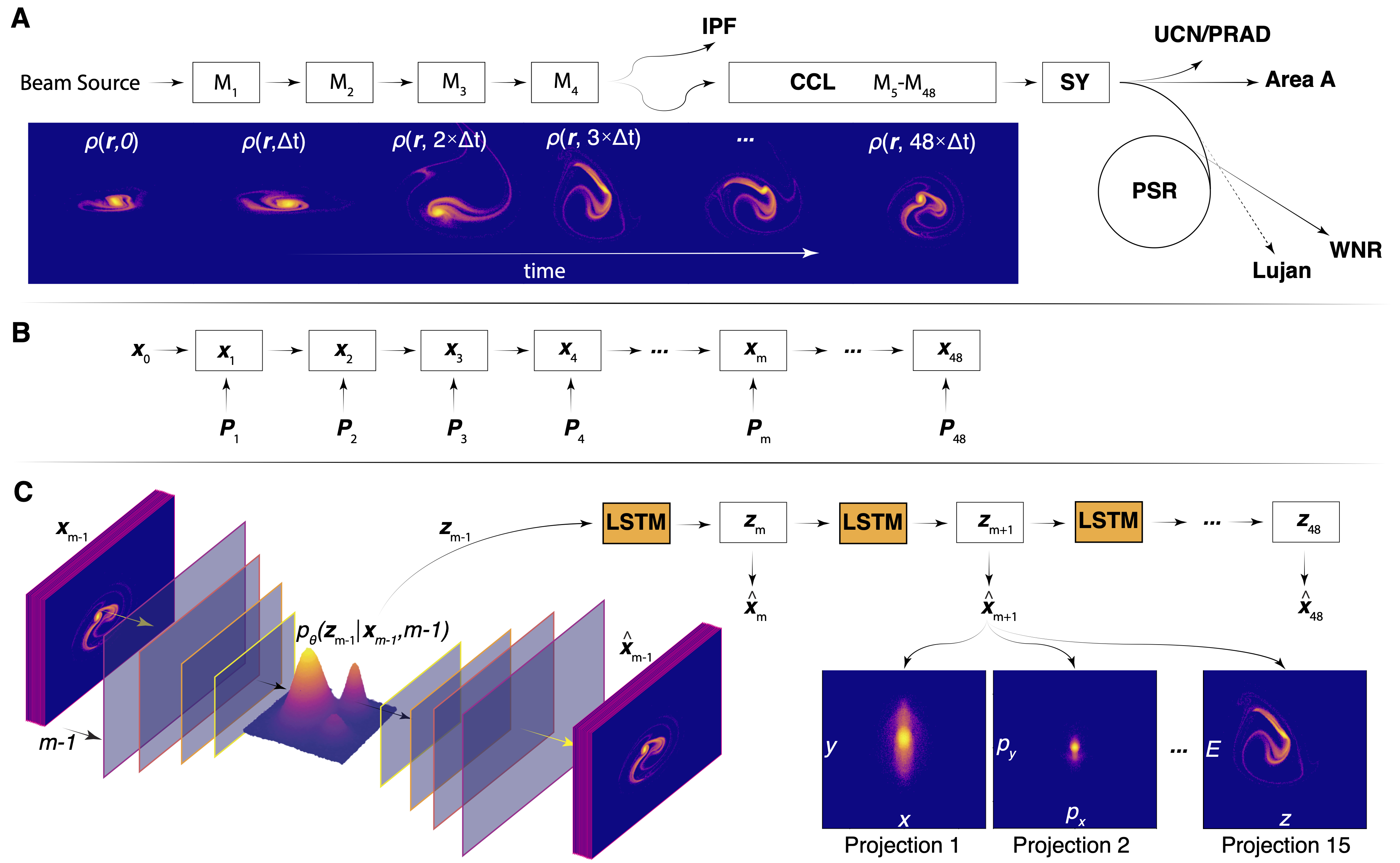}
    \caption{(\textbf{A}) A layout of the LANSCE linear accelerator and the related scientific areas is shown together with the 2D $(z,E)$ projection of the beam's charge density $\rho$ at various times along the accelerator. Modules M$_1$–M$_{48}$ are the resonant structures which accelerate the charged particle beam up to a final energy of 800 MeV. (\textbf{B}) Simplified representation of the beam where $\mathbf{x}_m$ represents all 15 of the 2D projections of the beam's 6D phase space when passing through M$_{m}$. $\mathbf{P}_m$ represents the amplitude and phase set-point of the electromagnetic field in M$_{m}$. (\textbf{C}) The inputs to the CVAE are $(\mathbf{x}_m,m)$, all 15 projections and also we condition the encoder on module number $m$ which is equivalent to conditioning on the amount of time the beam has spent moving through the accelerator. The CVAE allows us to study the beam dynamics based on a very low-dimensional latent representation of the beam's phase space. The beam's phase space $\mathbf{x}_{m-1}$ is embedded into a latent vector $\mathbf{z}_{m-1}$, which is then directly mapped to a latent vector representing the beam's subsequent state $\mathbf{z}_{m}$ by the LSTM, from which the decoder of the CVAE generates an estimate $\hat{\mathbf{x}}_{m}$ of the beam's phase space (only the $(x,y)$, $(p_x,p_y)$ and $(z,E)$ projections shown here).}
    \label{fig:summary_figure}
\end{figure}

Recently, latent space evolution-based models have gained attention in learning spatiotemporal dynamics. In these models, a dimensionality reduction framework (principal component analysis or a nonlinear autoencoder) is used to learn spatial correlations by mapping higher dimensional images to lower dimensional latent space. Such approaches increase computational efficiency by working in a lower dimensional feature space. For example, in Ref.~\cite{scheinker2021adaptive}, an autoencoder-based convolutional neural network maps high dimensional images representing the states of a complex time-varying system to a low-dimensional latent representation from which subsequent states of the system are then reconstructed. It was shown that by applying adaptive feedback directly within the latent representation an unknown time-varying system's properties could be quickly tracked with time. Latent space evolution models have also been applied for learning phase-field-based microstructure evolution, utilizing joint two-point statistics and principal component analysis for spatial correlations, where an LSTM was employed to predict the microstructure evolution of a two-phase mixture \cite{montes2021accelerating}. The combination of autoencoders and LSTMs is also employed for addressing fluid flow problems \cite{wiewel2019latent, nakamura2021convolutional, maulik2021reduced, vlachas2022multiscale}.

The latent embedding-based approach combined with LSTM has demonstrated potential benefits, showcasing increased speed when compared to ConvLSTM and DCGAN approaches \cite{montes2021accelerating}. Additionally, it has been observed that an encoder-decoder network extracts more representative lower-dimensional features compared to an encoder network in ConvLSTM \cite{rautela2022delamination}. The performance of a recurrent neural network trained to learn long temporal correlations is found to be superior in the presence of such features \cite{montes2021accelerating}. In the latent representations learned by vanilla autoencoders, the inputs may be mapped to disjoint locations, and special care has to be taken in the form of additional regularization on the latent representation in order to guide the autoencoder towards developing dense embeddings which can be smoothly traversed for generative purposes, as was done in Refs.~\cite{scheinker2021adaptive} and \cite{scheinker2023adaptive}. Variational autoencoders (VAE) are probabilistic models that naturally lead to dense latent space representations that can be smoothly traversed, making them well suited for probabilistic density estimation \cite{kingma2013auto}. Furthermore, a trained VAE can generate new realistic samples, similar to regularized autoencoders and generative adversarial networks, with the added benefit of explicitly learning a probability distribution over input data in the latent space \cite{hartmann2022unsupervised}. In this paper, we introduce a Conditional Latent Autoregressive Recurrent Model (CLARM) where we used a Conditional Variational Autoencoder (CVAE) to learn spatial correlations in a lower-dimensional latent space. Then, we utilize a recurrent neural network, such as LSTM, to predict future temporal states autoregressively based on limited previous states in the latent space. We apply CLARM to generate and forecast charged particle beam dynamics in linear accelerators, specifically implementing the algorithm for a charged particle beam in the linear accelerator at the Los Alamos Neutron Science Center (LANSCE) at Los Alamos National Laboratory (LANL).


\subsection{Spatiotemporal dynamics of charged particles}
In a linear accelerator, the  dynamics of charged particles evolve temporally in a six-dimensional phase space made up of three position and three momentum components for each particle i.e., $(x, y, z, p_x, p_y, p_z)$ with $z$ typically chosen as the direction along the accelerator axis \cite{scheinker20236d}. An ensemble of billions of particles is typically described in terms of a time-varying density function $\rho(x,y,z,x',y',E,t)$ where $(x',y') = (p_x/p_z,p_y/p_z)$ are the angles at which the particles are moving relative to the accelerator axis. The dynamics of a beam's phase space density function are governed by the relativistic Vlasov equation
\begin{eqnarray}
    \frac{\partial \rho}{\partial t} &=& -\mathbf{v}\cdot \nabla_{\mathbf{x}}\rho + \frac{\partial\mathbf{p}}{\partial t}\cdot\nabla_{\mathbf{p}}\rho, \label{beam_dynamics1} \\
    \frac{\partial\mathbf{p}}{\partial t} &=& q\left ( \mathbf{E}(x,y,z,t) + \mathbf{v}\times \mathbf{B}(x,y,z,t)\right ), \label{beam_dynamics2}
\end{eqnarray}
with $\mathbf{v}$ and $\mathbf{p}$ representing velocity and relativistic momentum:
\begin{eqnarray}
    \mathbf{v} &=& \left( \frac{dx}{dt},\frac{dy}{dt},\frac{dz}{dt} \right ), \quad \mathbf{p} = (p_x,p_y,p_z) = \gamma m \mathbf{v}, \nonumber \\
    \gamma &=& \frac{1}{\sqrt{1-v^2/c^2}}, \quad v=|\mathbf{v}|,
\end{eqnarray}
where $c$ is the speed of light. The electromagnetic fields, $\mathbf{E}$ and $\mathbf{B}$ each have external and beam-based sources. The beam-based sources are the electromagnetic fields of the beams themselves, as defined by Maxwell's equations, which act back on the particles via space charge forces and distort the 6D phase space. The external sources are the electromagnetic devices such as resonant accelerating structures and magnets throughout a particle accelerator which accelerate and focus the beam. Relative to Equation \ref{spatiotemporal} above, the terms $\mathbf{P}$ are represented by the external sources of the electromagnetic fields $\mathbf{E}$ and $\mathbf{B}$. At large accelerators such as LANSCE, where the beam is very intense, the powerful self fields of the beam distort and filament the beam's phase space via nonlinear space charge forces resulting in highly complex phase space distributions, as shown in Figure \ref{fig:summary_figure}. 

In order to precisely control and optimize beam parameters it would be very useful to have a clear view of the intricate details of a beam's 6D phase space. Unfortunately, measuring a beam's 6D phase space has so far been impossible. An iterative method was recently demonstrated in which various low dimensional projections of $\sim$5.7 million individual beams were measured over 32 hours and then assumed to vary very little from shot-to-shot so that, when considered as identical beams, a full 6D estimate of the phase space could be reconstructed \cite{cathey2018first}. Fast, single shot detailed beam measurements that are available are destructive in nature and provide 2D projections of the beam's 6D phase space distribution. For example, by passing a beam through a scintillating target followed by an optical detector the 2D $(x,y)$ projection of the beam can be measured. Another approach is to streak the beam with a high frequency electromagnetic field before passing it through a strong magnet that separates the beam as a function of energy, before impacting a scintillator in order to extract a 2D $(z,E)$ projection. 

Unfortunately such methods are typically only used for electron beams, and are much more difficult for high intensity proton beams which will destroy such measurement devices. Additionally there are other practical challenges that limit measurements such as resolution limits with short duration beam pulses and short run times which make meaningful data gathering impractical. Finally, there is also the problem of distribution shift due to the fact that accelerator components and the initial beam distribution change with time.

Therefore, while particle accelerators are high-dimensional time-varying systems governed by hundred to thousands of parameters and have very complex beams, detailed non-invasive phase space diagnostics are incredibly limited. Physics-based simulations are commonly used to understand and investigate beam diagnostics across different modules of linear accelerators. However, the available simulators \cite{tenenbaum2005lucretia,young2003particle,pang2014gpu} for simulating the 6D phase space of charged particle beams are computationally expensive and require domain knowledge with expert supervision. Also, the simulations cannot capture distribution shifts in the system parameters and uncertainties in the phase space of charged particles without an uncertainty propagation framework \cite{adelmann2019nonintrusive}. This limitation restricts their applicability for inverse problems like parameter estimation, system identification, and control \cite{newton1970inverse}. It is imperative to develop computationally efficient and robust probabilistic surrogate models to beam dynamics accurately.

\subsection{Review of deep learning for accelerators}
Deep learning is now regularly used for particle accelerator applications. In Ref.~\cite{scheinker2018demonstration}, a first demonstration of adaptive ML was performed combining adaptive feedback with deep neural networks for automatic shaping of the $(z,E)$ longitudinal phase space of the LCLS Free Electron Laser’s electron beam. In Ref.~\cite{wolski2022transverse}, a method was developed for 4D transverse phase space tomography. In Ref.~\cite{mayet2022predicting}, a  method was developed for predicting the transverse emittance of space charge-dominated beams and demonstrated. In Ref.~\cite{zhu2021high}, virtual diagnostics were developed for accelerator injector tuning. In Ref.~\cite{emma2018machine} a virtual diagnostic was developed. Deep Lie map networks were developed to identify magnetic field errors based on beam position monitor measurements in synchrotrons \cite{caliari2023identification}. In Ref.~\cite{breckwoldt2023machine}, Bayesian optimization was used for calibration of intense x-ray FEL pulses. 

In Ref.~\cite{ivanov2020physics}, polynomial neural networks are developed with symplectic regularization to represent Taylor maps of particle dynamics. In Ref.~\cite{meier2022optimizing}, deep reinforcement learning was used to optimize the design of a superconducting radio-frequency gun. In Ref.~\cite{obermair2022explainable}, interpretable ML-based models were developed for predicting breakdowns in high-gradient cavity data. In Ref.~\cite{tennant2020superconducting}, DL is used to classify superconducting radio-frequency (SRF) cavity faults.  In Ref. \cite{li2018genetic}, a genetic algorithm enhanced by ML is used for multi-objective optimization for the dynamic aperture a storage ring. In Ref.~\cite{cropp2023virtual}, ML methods have been combined with multilinear regression to create virtual time of arrival and beam energy diagnostics at HiRES. While some of the methods described above provide surrogate models for specific aspects of accelerator dynamics, none of them deliver a detailed view of an accelerator beam's evolving phase space throughout an entire machine. 

In Ref.~\cite{scheinker2021adaptive}, a general-purpose algorithm was developed that combines model independent adaptive feedback \cite{scheinker2016bounded} and DL with particle tracking codes to efficiently reconstruct phase space distributions. This method was focused on making predictions at specific isolated accelerator locations without incorporating the temporal evolution of the beam from one location to another. In Ref.~\cite{scheinker2023adaptive}, the initial conditions of a beam, conditioned on beam current and accelerator magnet strength, were encoded into a low-dimensional (2D) latent space representation, from which the 15 unique 2D projections of the beam's 6D phase space downstream were generated. Adaptive feedback within the latent representation automatically tracked a time-varying beam. This method was focused on an input-output mapping from the entrance to the exit of the accelerator and did not study the temporal evolution of the beam between the two end points.

\subsection{Summary of main results}
The existing state-of-the-art latent space evolution models for solving spatiotemporal dynamics across diverse areas of physics face a limitation due to the absence of specialized probabilistic models, which could be leveraged for generation purposes. The existing accelerator literature does not explore deep learning methods for modeling charged particle dynamics as a spatiotemporal phenomenon. This lack of a comprehensive modeling framework for charged particle dynamics forces the results to focus on specific accelerator locations without a mechanism for the exploitation of forecasting capabilities in other locations.

Fig. \ref{fig:summary_figure}(B) depicts a simplistic mathematical representation of Eq. \ref{spatiotemporal} for the linear accelerator with M modules. The beam, dynamics described by Eqs. \ref{beam_dynamics1}, \ref{beam_dynamics2} are a special case of Eq. \ref{spatiotemporal}, as they describe the evolution of a charged particle beam's phase space density $\rho$. We consider the evolution of these dynamics over a finite grid of space and time points $t\in\{0,\Delta_t,2\times\Delta_t,\dots,M\times\Delta_t\}$. For notational simplicity we refer to $\rho$ at these times as $\{\rho_1,\rho_2,\dots,\rho_M\}$. 
Since we are working with the phase space of charged particles denoted as $X$ rather than phase space density $\rho$, we adopt notations using $X$ instead of $\rho$. Our approach is motivated by the fact that such a discretized spatiotemporal problem can be solved by an autoregressive model where the current state of $X$ depends only on previous states, $X_{m} = H(X_1, X_2,..., X_{m-1})$, where $H$ is a unknown nonlinear function. As mentioned earlier, the RF parameters of accelerators have a time-varying nature. The beam's phase space is manipulated by the RF parameters, and these uncertainties propagate and appear in the phase space of the charged particles. Therefore, the parameters $P_1, P_2, ... , P_M$ and phase space $X_1, X_2,..., X_M$ shown in Fig.~\ref{fig:summary_figure}(B) have uncertainties due to the time-varying nature of the system. Such intricate problems can be effectively learned through a probabilistic approach, requiring the calculation of the joint probability distribution $P(X_1, X_2, ... ,X_M)$. One way to compute this distribution is the autoregressive solution, where $P(X_1, X_2, ... ,X_M)$ is factored as $P(X_1)P(X_2|X_1)P(X_3|X_1,X_2,X_3)$...$P(X_M|X_1,X_2,...,X_{M-1})$. In the context of linear accelerators, $X$ is a high-dimensional object comprising positions and momentum of billions of particles. Learning such a high-dimensional probability distribution and calculating $P(X_{m} | X_1, X_2, ..., X_{m-1})$ to estimate $X_{m}$ at any given module is computationally infeasible. Fortunately, latent variable models (like variational autoencoders) transform the higher-dimensional distribution $P(X_1, X_2, ... ,X_M)$ into a lower-dimensional distribution $P(\mathbf{z}_1, \mathbf{z}_2, ..., \mathbf{z}_{M})$ using Bayesian inference \cite{kingma2013auto}. The autoregressive approach can then be applied to this lower-dimensional distribution, calculating $P(\mathbf{z}_{m} | \mathbf{z}_1, \mathbf{z}_2, ..., \mathbf{z}_{m-1})$ to estimate $\mathbf{z}_{m}$, as shown in Fig. \ref{fig:summary_figure}(C).

Current methods, such as a linear autoregression model, using $\mathbf{z}_t = w_0 + \sum^{t-1}_{i=1} w_i \mathbf{z}_i$, can be employed to learn the temporal dynamics in the latent space. But, its linearity limits its ability to learn long temporal correlations \cite{nelson1998time}. Non-linear models like Masked Autoencoder Density Estimation (MADE) \cite{germain2015made} and Neural Autoregressive Distribution Estimation (NADE) \cite{uria2016neural} have also been utilized to learn the autoregressive nature using neural networks, but they can only be applied for single-dimensional $z$. While a simple recurrent neural network (RNNs) lacks the capability to learn long-term correlations, an LSTM is designed to tackle this problem. The LSTM needs to be trained to consider the autoregressive nature of the latent space and learn $\mathbf{z}_{m} = G(\mathbf{z}_{1:m-1})$ for all $m$, where $G$ is combined functional relationships of LSTM \cite{toneva2022combining}. This can be accomplished by training the LSTM with a variable input size. Once the temporal dynamics are learned in the latent space, the decoder part of the VAE transforms it back to the high-dimensional space, obtaining $P(X_{m} | X_1, X_2, ..., X_{m-1})$, as shown in Fig. \ref{fig:summary_figure}(C).

As described above, in this paper, we introduce a novel deep learning modeling framework, wherein we study the full spatiotemporal dynamical nature of the evolution of a charged particle beam through various sections of a linear accelerator. The dynamics of the beam depend on accelerator parameter settings, such as resonant accelerating structures that generate the electric fields in Equation \ref{spatiotemporal}. We work directly with images which are all 15 unique 2D projections of the beam's 6D phase space at 48 different accelerator locations. We take advantage of the fact that information regarding parameter settings is encoded within the intricate phase space of the charged particle beam as it traverses various accelerating sections. This gives us the ability to eliminate the RF field settings as conditional inputs so that we train the deep learning model on the images directly. We refer to our approach as a Conditional Latent Autoregressive Recurrent Model (CLARM) designed to capture the evolution of charged particle beams in unsupervised settings. Our proposed method involves a two-step deep learning approach: 1.) utilizing a conditional variational autoencoder (CVAE) to learn a low-dimensional latent space distribution (8D) of the high dimensional (15 images at 256$\times$256 pixels each is a $\sim10^6$ dimensional object) phase space of charged particles, and 2.) employing a Long Short-Term Memory (LSTM), a type of recurrent neural network, to learn the temporal dynamics within the latent space. The LSTM is trained to forecast beam behavior at subsequent accelerator locations based on previous locations in an autoregressive setting. By integrating these two neural network structures into a unified architecture within an autoregressive framework, CLARM effectively learns spatial behavior using CVAE and temporal dynamics using LSTM. It gives the model two promising abilities, allowing both the generation of realistic projections across different modules and the forecasting of phase space in further modules. 

The probabilistic nature of the model can also provide an additional benefit, facilitating the analysis of uncertainties \cite{acharya2023learning}. The proposed model addresses challenges faced by existing models in learning spatiotemporal dynamics, such as expensive computations in ConvLSTM \cite{shi2015convolutional} and DCGAN \cite{cheng2020data}, mixed spatial features and temporal dynamics in 3DCNN/4DCNN \cite{wandel2021teaching}, computational complexity, lack of scalability, and robustness issues in GNNs \cite{wu2020comprehensive}, linear representation of data in the latent space through PCA \cite{montes2021accelerating}, and explainability concerns in the latent space with non-linear dimensionality reducers like autoencoders \cite{nakamura2021convolutional}.

The key contributions of this paper are summarized as:
\begin{itemize}
    \item Proposing an unconventional approach to model the evolution of charged particles in an accelerator as a spatiotemporal dynamical phenomenon.
    \item Introducing CVAE, in contrast to the AE used in the literature, provides the ability to generate realistic phase space projections across different modules of the linear accelerator.
    \item Integrating CVAE and LSTM to independently learn spatial and temporal dynamics and combining them in an autoregressive loop.
    \item End-to-end forecasting of the future behavior of charged particles given previous behaviors autoregressively.
    \item Providing interpretability and explainability through latent space visualization and lower-dimensional principal component analysis (PCA), t-stochastic nearest neighbors (t-SNE), and uniform manifold approximation and projections (UMAP).
\end{itemize}

The paper is organized as follows: Section~\ref{sec:methods} gives details about the multi-particle tracking model and CLARM, Section~\ref{sec:results} presents results and discussions on the reconstruction ability, latent space visualization, generative and forecasting ability of the CLARM. The paper is concluded in Section~\ref{sec: conclusions}.


\section{Methods} \label{sec:methods}
\subsection{Multi-particle tracking simulations} \label{ssec:simulations}
\begin{figure}[hbt!]
    \centering
    \begin{minipage}[b]{1.0\linewidth}
        \centering
        \includegraphics[width=1.0\textwidth]{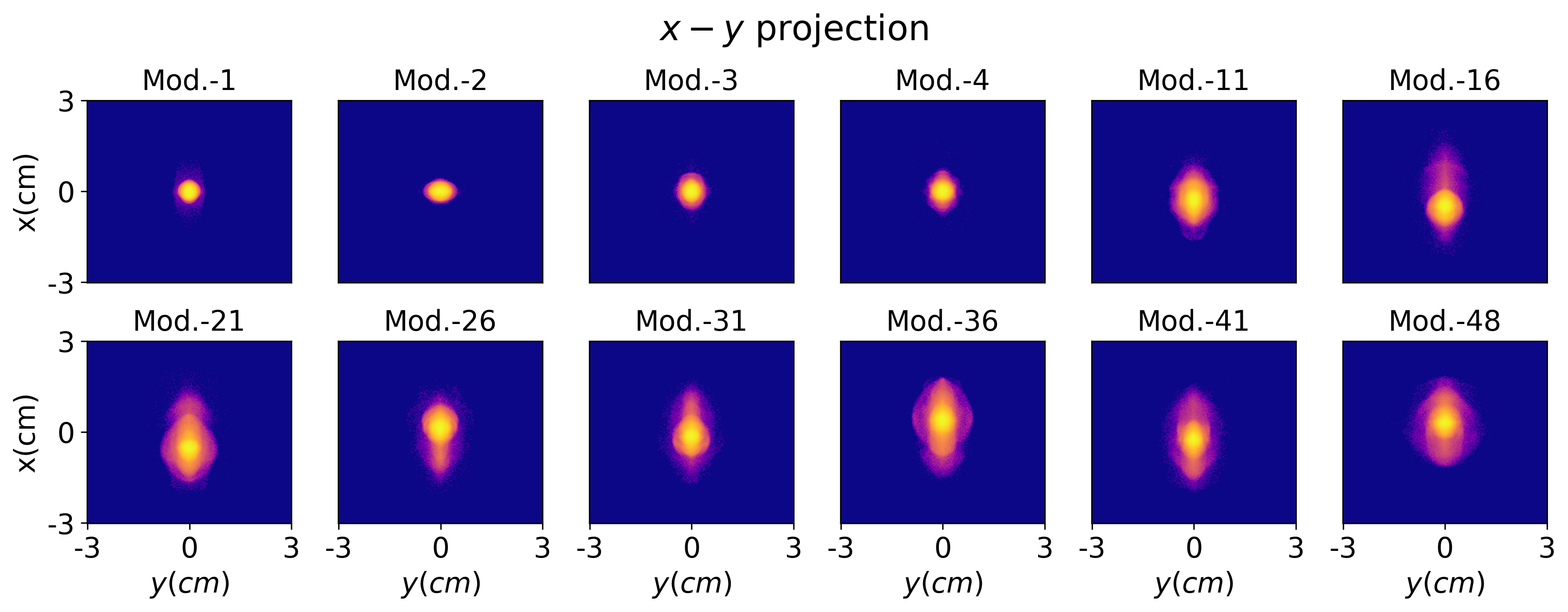}
    \end{minipage}
    \begin{minipage}[b]{1.0\linewidth}
        \centering
        \includegraphics[width=1.0\textwidth]{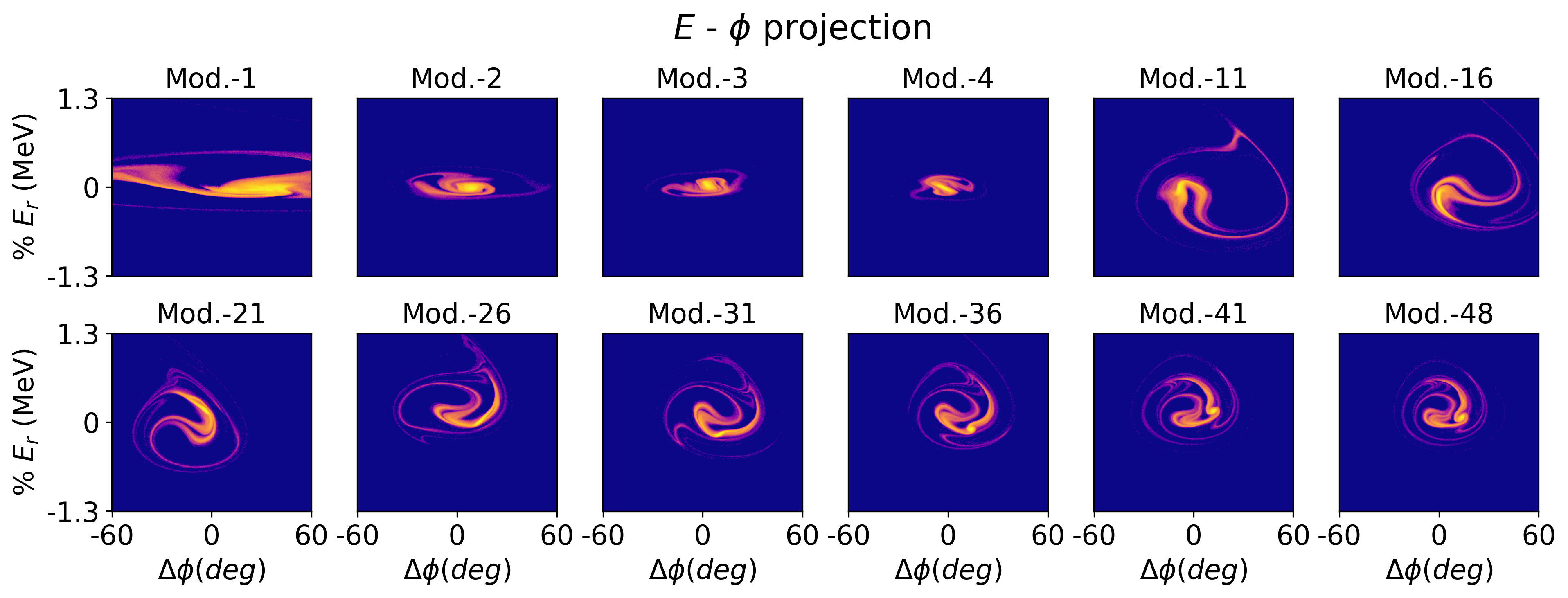}
    \end{minipage}
    \begin{minipage}[b]{1.0\linewidth}
        \centering
        \includegraphics[width=1.0\textwidth]{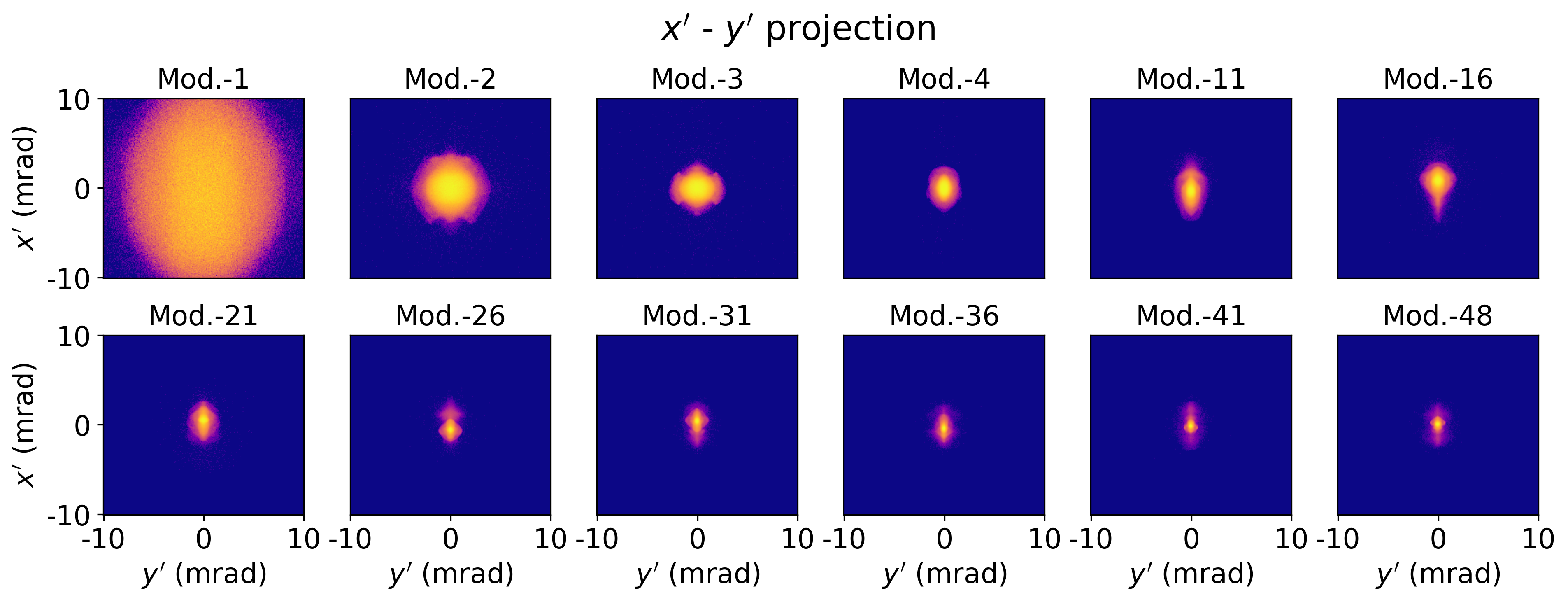}
    \end{minipage}
    \caption{Three out of fifteen projections ($x-y$, $E-\phi$, $x'-y'$) of the 6D phase space of charged particle beams at different modules. The plots are shown on a logarithmic scale for better visualization. The plots illustrates how the projections evolve as they experience RF electromagnetic fields while moving through different modules.}
    \label{fig:dataset}
\end{figure}

In this study, we focus on the LANSCE linear accelerator at Los Alamos National Laboratory \cite{wangler2008rf}. The LANSCE accelerator provides high-intensity $H^+$ and $H^-$ particle beams up to 800 MeV for various scientific experiments. The LANSCE accelerator includes H+ and H- injectors, each featuring an ion source and a Cockcroft–Walton accelerator, to produce a 750 KeV beam. A beam transport system, approximately 12 meters long and equipped with magnets and RF cavities, focuses, guides, and bunches the beams. The beams are then injected into the 201.25-MHz Drift Tube linac (DTL) for acceleration to 100 MeV. The DTL, divided into four tanks or modules, each powered by a separate RF amplifier, and contains numerous focusing quadrupole magnets. The beam then enters the 805 MHz coupled-cavity linac (CCL) to accelerate the beams to 800 MeV, which has an additional 44 modules, each being an independently driven accelerating structure, and quadrupole doublets along its 726-m length. Fig.~\ref{fig:dataset}(a) presents a simplified schematic of the LANSCE linear accelerator. Accelerating modules $M_1-M_4$ belong to the 201.25 MHz DTL section. After the $M_4$ module, a 100 MeV beam is used for the isotope production facility (IPF). Modules $M_5-M_{48}$ are located in the 805 MHz CCL section where the beam reaches a kinetic energy of 800 MeV. The beam is then used for applications in areas such as ultracold neutron source (UCN) studies, proton radiography (PRAD) for dynamic material imaging, and high-energy physics experiments. Additionally, beams can be directed to the proton storage ring (PSR) for creating highly intense beam bunches, used in the weapons neutron research (WNR) facility or the Lujan center for neutron scattering \cite{scheinker2021extremum}.

During the operation of the accelerator, system parameters such as magnet and RF field set points are adjusted manually to achieve minimal beam loss. This process is laborious and time consuming and often falls short of identifying the optimal performance. In order to achieve optimal performance, insights into beam dynamics play an important role. To this end, several simulations tools have been developed \cite{tenenbaum2005lucretia,young2003particle,pang2014gpu}. Notably, HPSim, an advanced, open-source tool developed at LANL, enables rapid, online simulations of multiple-particle beam dynamics \cite{pang2014gpu}. HPSim solves Vlasov-Maxwell equations to calculate the effects of external accelerating and focusing forces on the charged particle beam as well as space charge forces within the beam. In this paper, we utilize HPSim to collect limited data for the DL model. The simulations are performed on macroparticles (1,048,576 in number). HPSim contains a realistic representation of the true beam used at LANSCE, able to simulate the beam as it passes through 48 modules of the DTL and the CCL. Each module comprises two RF set points, one dictating the phase of the electromagnetic wave responsible for beam acceleration, and the other determining the amplitude.

To generate the dataset from HPSim, the RF set points (amplitude and phase) for the first four modules are randomly sampled from a uniform distribution keeping the rest of the set points of 44 modules at a mean value. Other beam and accelerator parameters, like the initial beam condition, are also set to constant realistic values. Using the RF set points as inputs to the simulation, HPSim provides a six-dimensional phase space of the charged particle beam in the form of 15 unique projections ($x-p_x$, $x-y$, $x-p_y$, $x-z$, $x-p_z$, $y-p_x$, $y-p_y$, $y-z$, $y-p_z$, $z-p_x$, $z-p_y$, $z-p_z$ ($E-\phi$), $p_x-p_y$, $p_x-p_z$, and $p_z-p_y$) at each of the 48 modules. 1400 different simulations are performed to collect training data and another 100 simulations for the test data. In Fig.~\ref{fig:dataset}, 3 out of 15 phase space projections ($x-y$, $E-\phi$, and $x'-y'$) are plotted for different accelerating modules. The plots are transformed on a logarithmic scale to squeeze the broader range of intensity levels of the pixels into a more compact range. This aids in enhancing the clarity of the plots for visualization purposes. The original untransformed projections, which are also utilized for training the DL models, are presented in supplementary Fig.~1. It can be observed that the projections evolve as they interact with RF electromagnetic fields while traversing through various modules. It is also evident that the projections exhibit greater variations from module to module when observing earlier modules, with the variation gradually reducing as we progress towards the later modules. This can be observed clearly with the $E-\phi$ projection where the images appear to be rotated versions of each other in the later modules. This characteristics of the projections has implications while generating new projections across different modules, which will be discussed in the later section.

\subsection{Conditional latent autoregressive recurrent model} \label{ssec:theory}
\begin{figure}[hbt!]
\centering
\includegraphics[width=1.0\textwidth]{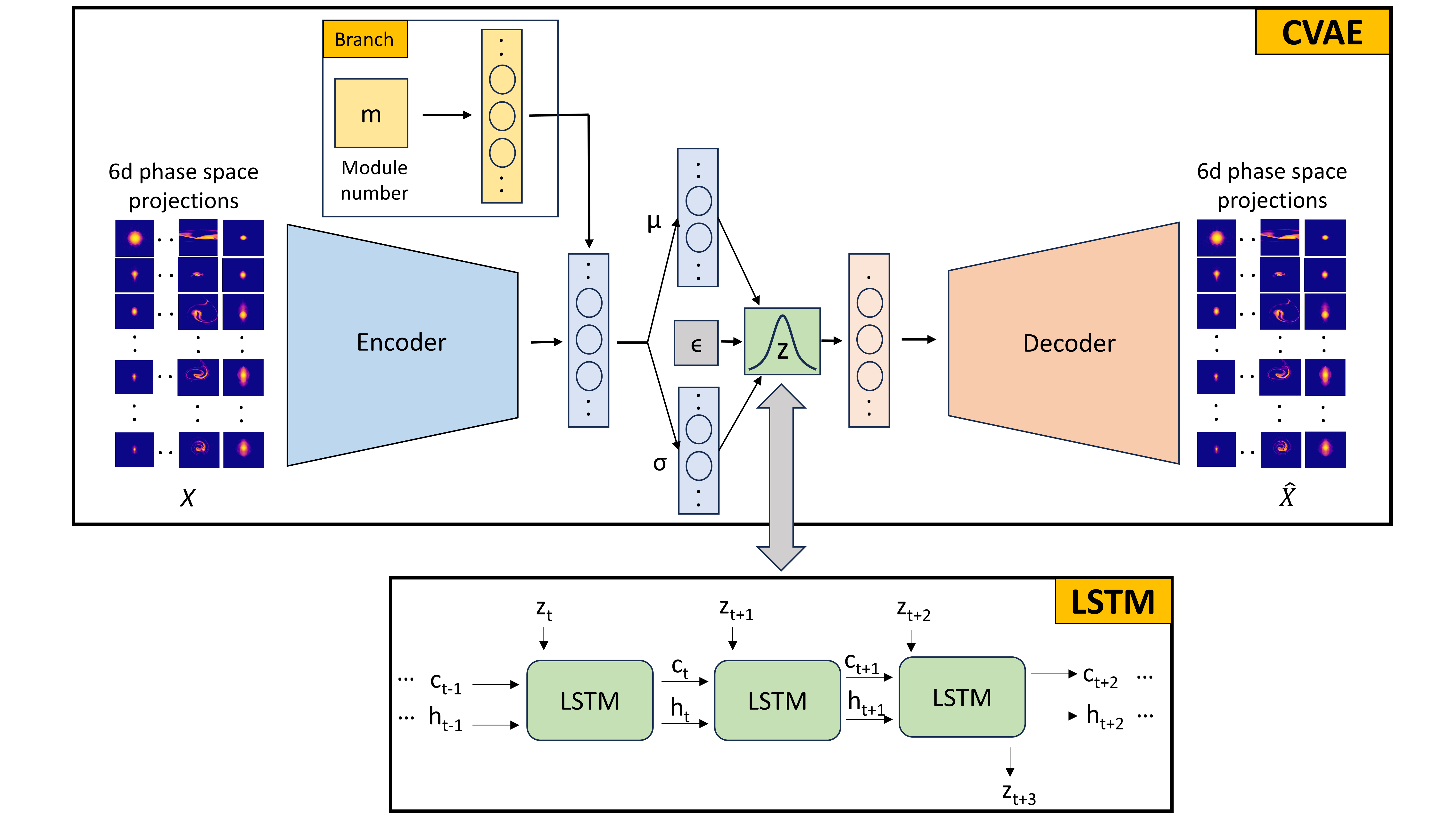}
\caption{CLARM is a two step DL framework which combines a CVAE and an LSTM acting on the latent embedding of the CVAE. The CVAE transforms the phase space projections ($\mathbf{x}$) into the latent space ($\mathbf{z}$) and learns a probabilistic low dimensional distribution of ($\mathbf{x}$). The LSTM learns the temporal dynamics in the latent space. The CVAE and LSTM are coupled together in an autoregressive loop for forecasting phase space projections through the system.}
\label{fig:CLARM}
\end{figure}

CLARM leverages a Conditional Variational Autoencoder (CVAE) to transform the 15 unique projections of the 6D phase space of charged particle beams into a lower-dimensional latent space distribution. Subsequently, a Long Short-Term Memory (LSTM) network learns the temporal dynamics within this latent space. Both networks are integrated into an autoregressive framework. The uniqueness of CLARM lies in its ability to independently learn spatial and temporal dynamics through a two-step process. In simpler terms, learning spatial dynamics involves representing the 6D phase space of charged particles in the latent space, while learning temporal dynamics entails creating temporal structure in the latent space as charged particles (in the form of extracted features) move through different modules. The architecture of the CLARM is represented in Fig.~\ref{fig:CLARM}. The initial 15 unique phase space projections are depicted as 15 channels, each represented by a 256 $\times$ 256 pixel image, resulting in a $\sim10^6$ dimensional input that is encoded into an 8 dimensional latent space. The projections along with the module number (1-48) are input to the CVAE. The encoder performs convolution operations, and the extracted features are concatenated with the module number, and then are input to the CVAE's latent space. The learned latent space is then processed by an LSTM network, which is designed to forecast the next latent space point (corresponding to projections in the next module) based on previous points (corresponding to projections in the previous modules). The continuous latent space of the CVAE allows for conditional sampling, followed by a decoder which generates realistic projections across different modules of the accelerator.

A mathematical description of CLARM can be written in terms of the CVAE and LSTM, combined through latent space projections $\mathbf{z}$. The data distribution of $x$, parameterized by the neural network's parameters $\theta$ can be defined by the chain rule:
\begin{equation}
    p_{\theta}(\mathbf{x}) = \sum_{m=1}^{M} \int p_{\theta}(\mathbf{x}|\mathbf{z},m) p_{\theta}(\mathbf{z}|m)p_{\theta}(m)dz
\end{equation}
where the summation is over all the classes (module numbers, $m$) and M = 48.

The posterior distribution $p_{\theta}(\mathbf{z}|\mathbf{x},m)$ (conditioned on the the module number in our work) can be written using Bayes rule:
\begin{equation}
    p_{\theta}(\mathbf{z}|\mathbf{x},m) = \frac{p_{\theta}(\mathbf{x}|m,\mathbf{z})p_{\theta}(\mathbf{z}|m) p_{\theta}(m)}{p_{\theta}(\mathbf{x})}.
\end{equation}

The above posterior is intractable and cannot be solved. VAE introduces a parameterized distribution $q_{\phi}(\mathbf{z}|\mathbf{x},m)$ to reduce the dimensionality of the problem. It is an inference model that maps data $\mathbf{x}$ to the latent space $\mathbf{z}$ by approximating the posterior distribution $p_{\theta}(\mathbf{z}|\mathbf{x},m)$. Since, $p_{\theta}(m)$ is known and provides $p_{\theta}(\mathbf{z}|m)$, therefore, $p_{\theta}(\mathbf{x}|\mathbf{z},m) = p_{\theta}(\mathbf{x}|\mathbf{z})$. The loss function can be written in the form of Evidence Lower BOund (ELBO) \cite{sohn2015learning}:
\begin{eqnarray}
    EL(\theta,\phi;\mathbf{x}) &=& \mathbb{E}_{\mathbf{z} \sim q_{\phi}(\mathbf{z}|\mathbf{x}, m)} [\log p_{\theta}(\mathbf{x}|\mathbf{z})] \nonumber \\
    && - D_{KL}(q_{\phi}(\mathbf{z}|\mathbf{x},m)||p_{\theta}(\mathbf{z}|m)). \nonumber
\end{eqnarray}

In ELBO, the first term represents the reconstruction loss between the original and the predicted data whereas the second term is the KL divergence between the posterior distribution and the approximate posterior \cite{yin2021neural}. The second term is what differentiates VAEs from autoencoders. Autoencoders are not probabilistic models, they are trained only with reconstruction error and do not attempt to learn the probability density of the data. In vanilla autoencoders, if no additional regularization is applied on the latent space, they can map inputs to disjoint sets producing inaccurate estimates when moving through large empty regions of the latent space for sampling and generation \cite{notin2021improving}. One point to note here is that only the posterior distribution is conditioned on the module number $m$ because the module number is the input of the encoder only (Ref Fig.~\ref{fig:CLARM}). The set of equations changes depending upon the condition of the encoder, decoder or both \cite{lim2018molecular}. For the generation of realistic phase space projections in a particular module, the latent space is sampled randomly within the bounds of the latent space of the corresponding module and then decoded.

LSTM is a class of recurrent neural networks (RNNs) designed to process sequential data with long-term dependencies \cite{toneva2022combining}. LSTM presents a better alternative to simple RNNs and GRUs to learn the temporal dynamics due to their capacity to retain patterns learned earlier in the sequence and minimize the vanishing gradient problems \cite{karevan2020transductive}. In the context of this study, LSTM is employed as a forecasting model to predict future states (downstream modules) of the latent space based on limited initial states (upstream modules). The LSTM equations \cite{sak2014long}, originally expressed in terms of input $x_t$ and output $y_t$ at time $t$, are reformulated to forecast $\mathbf{z}_{m+1}$ (output) given $\mathbf{z}_m$ (input), with $m$ denoting the module number:
\begin{align}
    i_m = \sigma (W_{ii} \mathbf{z}_m + b_{ii} + W_{hi}h_{m-1} + b_{hi})\\
    f_m = \sigma (W_{if} \mathbf{z}_m + b_{if} + W_{hf}h_{m-1} + b_{hf})\\
    g_m = \tanh(W_{ig} \mathbf{z}_m + b_{ig} + W_{hg} h_{m-1} + b_{hg})\\
    o_m = \sigma (W_{io} \mathbf{z}_m + b_{io} + W_{ho} h_{m-1} + b_{ho})\\
    c_m = f_m \odot c_{m-1} + i_m \odot g_m\\
    h_m = o_m \odot \tanh(c_m) \\
    \mathbf{z}_{m+1} = \sigma (W_{hy} h_m + b_{hy})
\end{align}
where, $h_m$, $c_m$, $\mathbf{z}_m$ are the hidden state, cell state and input at module $m$, $h_{m-1}$ is the hidden state at module $m-1$. $i_m$, $f_m$, $g_m$, $o_m$ are respectively the input, forget gate, cell and output gates. $\sigma$ is the logistic sigmoid activation and $\odot$ is the Hadamard product. $W$ and $b$ are respective weights and bias matrices. A mean squared error-based loss function is used between true $\mathbf{z}_{m+1}$ and predicted $\hat{\mathbf{z}}_{m+1}$ values of latent point at module $m+1$. 
\begin{equation}
    MSE = L(\psi;\mathbf{z}) = \parallel \mathbf{z}_{m+1} - \hat{\mathbf{z}}_{m+1} \parallel_2
\end{equation}

The aforementioned equations are presented in terms of a single input and single output for simplicity. However, our LSTM network is designed with multiple inputs and a single output. Once both CVAE and LSTM are trained, they can be seamlessly integrated in an autoregressive loop to facilitate the forecasting of phase space projections across the entire accelerator. The forecasting methodology of CLARM is outlined in Algorithm~\ref{alg:predict}. The algorithm takes phase space projections across the initial modules (upstream locations) $X_{1:m}$ as input and the user-defined number of downstream modules to be forecasted, represented by $m_{end}$. Typically, $m_{end}$ is set to 48, corresponding to the index of the final module, but it can also be any other module number $> m_{start}$. In the algorithm, the encoder of CVAE projects $X_{1:m}$ into the latent space to obtain $\mathbf{z}_{1:m}$. The LSTM acts on the latent representation to forecast $\mathbf{z}_{m+1}$. The decoder of the CVAE then reconstructs $X_{m+1}$, providing the phase space projections at the succeeding module. The forecasted latent point $\mathbf{z}_{m+1}$ is concatenated with input $\mathbf{z}_{1:m}$ to form $\mathbf{z}_{1:m+1}$. This autoregressive procedure is iteratively applied to compute $\mathbf{z}_{m+2}$, followed by $X_{m+2}$, and so forth, until $m_{end}$.

\begin{algorithm}[h!]
\caption{Forecasting with CLARM}
\begin{algorithmic}[1]
    \State Input: $m_{start}$, $m_{end}$, trained CVAE, trained LSTM
    \For{$m$ = $m_{start}$ $\rightarrow$ $m_{end}$}
    \State $\mathbf{z}_{1:m}$ = CVAE-encoder($X_{1:m}$, $1:m$) 
    \State $\mathbf{z}_{m+1}$ = LSTM($\mathbf{z}_{1:m}$) 
    \State $X_{m+1}$ = CVAE-decoder($\mathbf{z}_{m+1}$)  
    \State $\mathbf{z}_{1:m+1}$ = concat($\mathbf{z}_{1:m}$, $\mathbf{z}_{m+1}$) 
    \State $\mathbf{z}_{1:m} \gets \mathbf{z}_{1:m+1}$
    \EndFor    
\end{algorithmic}
\label{alg:predict}
\end{algorithm}

CLARM is a two-step learning framework where the training is performed in two steps, however,  prediction remains a single-step process. The benefits of using a two-step learning framework over a single-step approach can be explained by considering factors such as training time and the learning process \cite{vlachas2022multiscale}. In the single-step framework, the LSTM must iterate over the latent space generated by CVAE at each epoch. It is observed that the training time needed to achieve a specific MSE loss value is comparatively longer in the single-step process. Additionally, a drawback emerges in the form of a noisy learning curve in the single-step process. This is due to the fact that while the CVAE is still learning, its latent representations are continuously changing and therefore the LSTM is attempting to learn the temporal dynamics of a wildly shifting distribution. If instead the CVAE is first trained until reconstruction accuracy is high, then the LSTM can be trained on the fixed learned latent distribution.


\section{Results}\label{sec:results}
\begin{figure}[hbt!]
    \centering
    \begin{minipage}[b]{0.48\linewidth}
        \centering
        \includegraphics[width=1.0\textwidth]{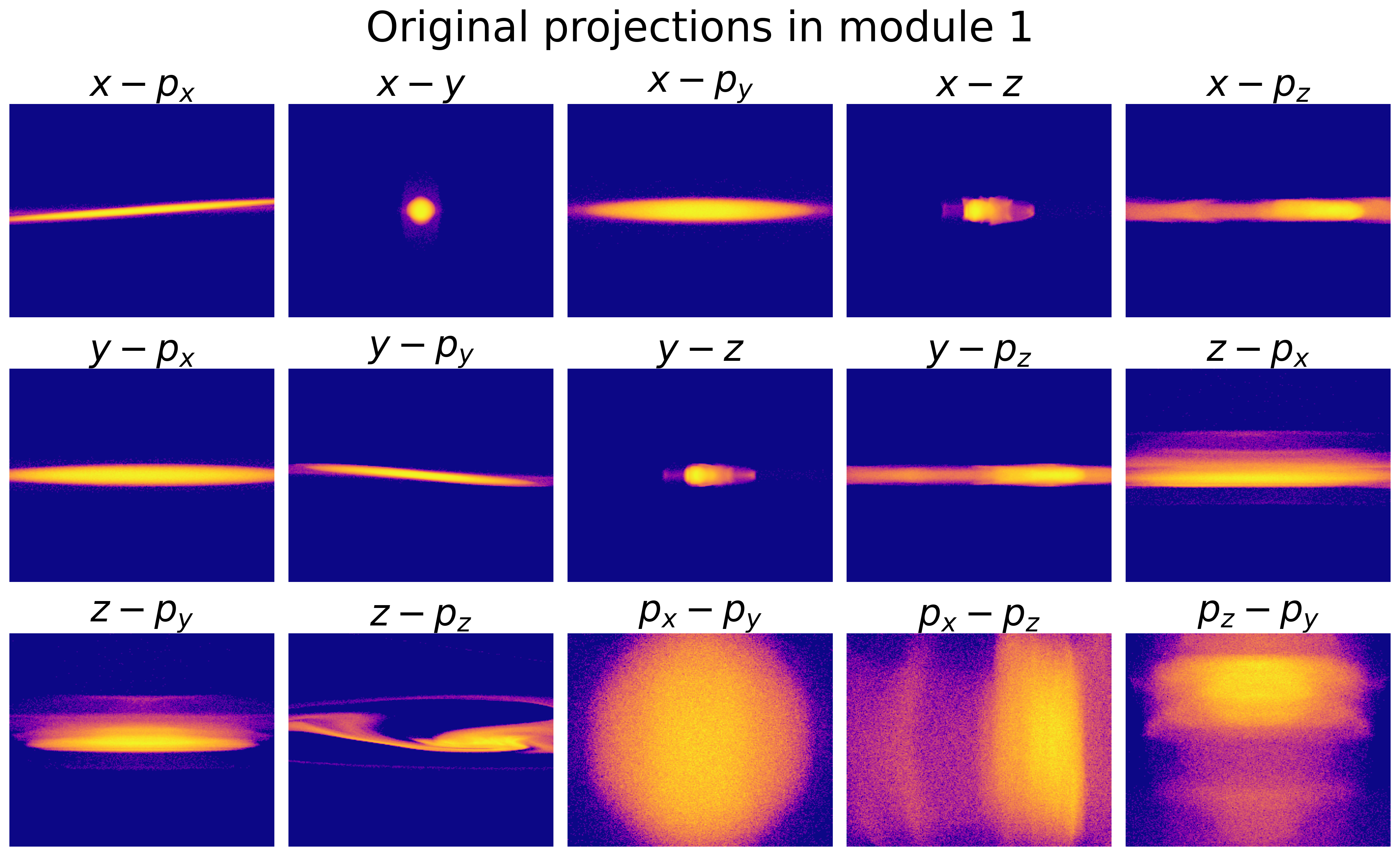}
        \vspace{1mm}
    \end{minipage}
    \begin{minipage}[b]{0.48\linewidth}
        \centering
        \includegraphics[width=1.0\textwidth]{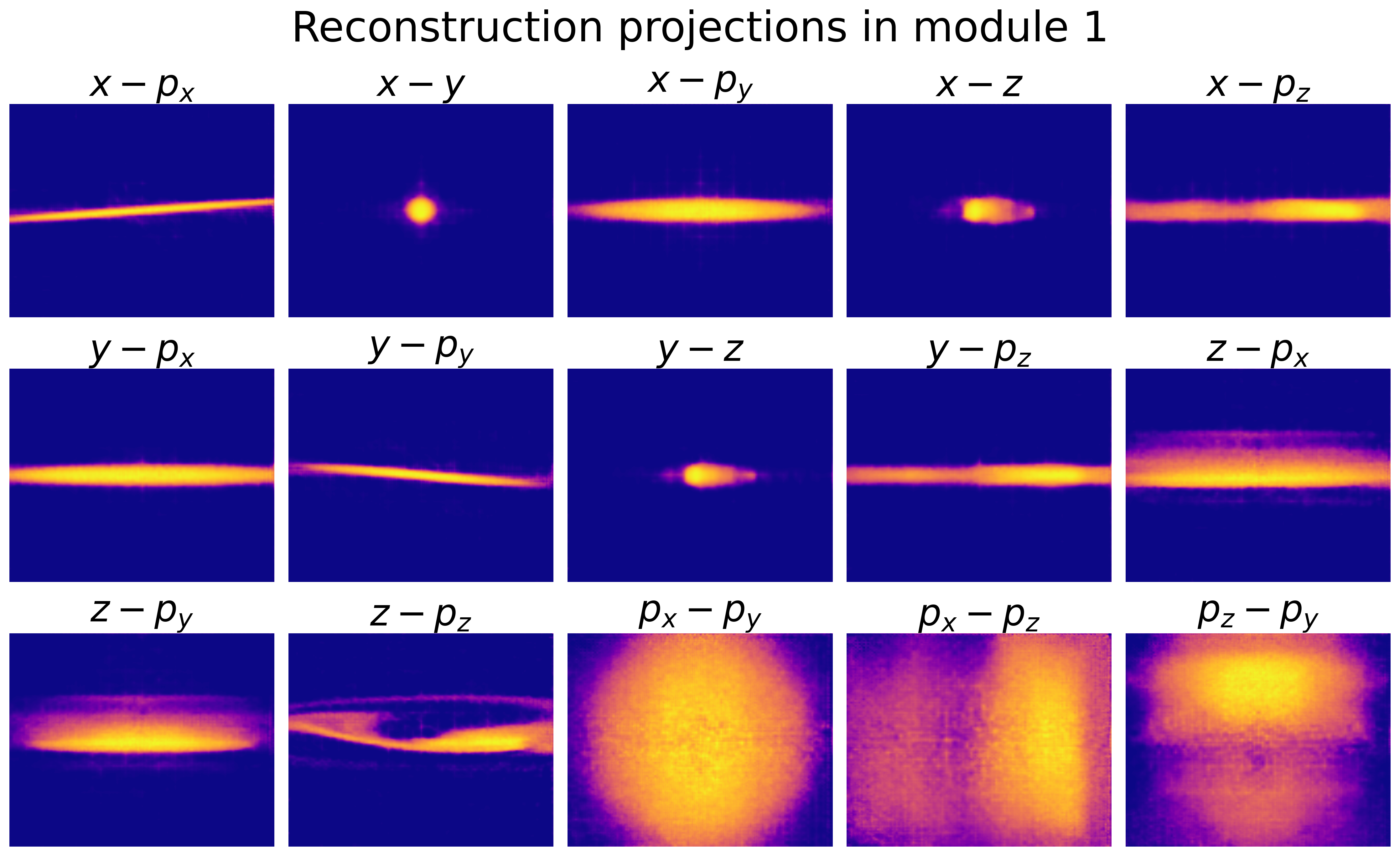}
        \vspace{1mm}
    \end{minipage}
    \begin{minipage}[b]{0.48\linewidth}
        \centering
        \includegraphics[width=1.0\textwidth]{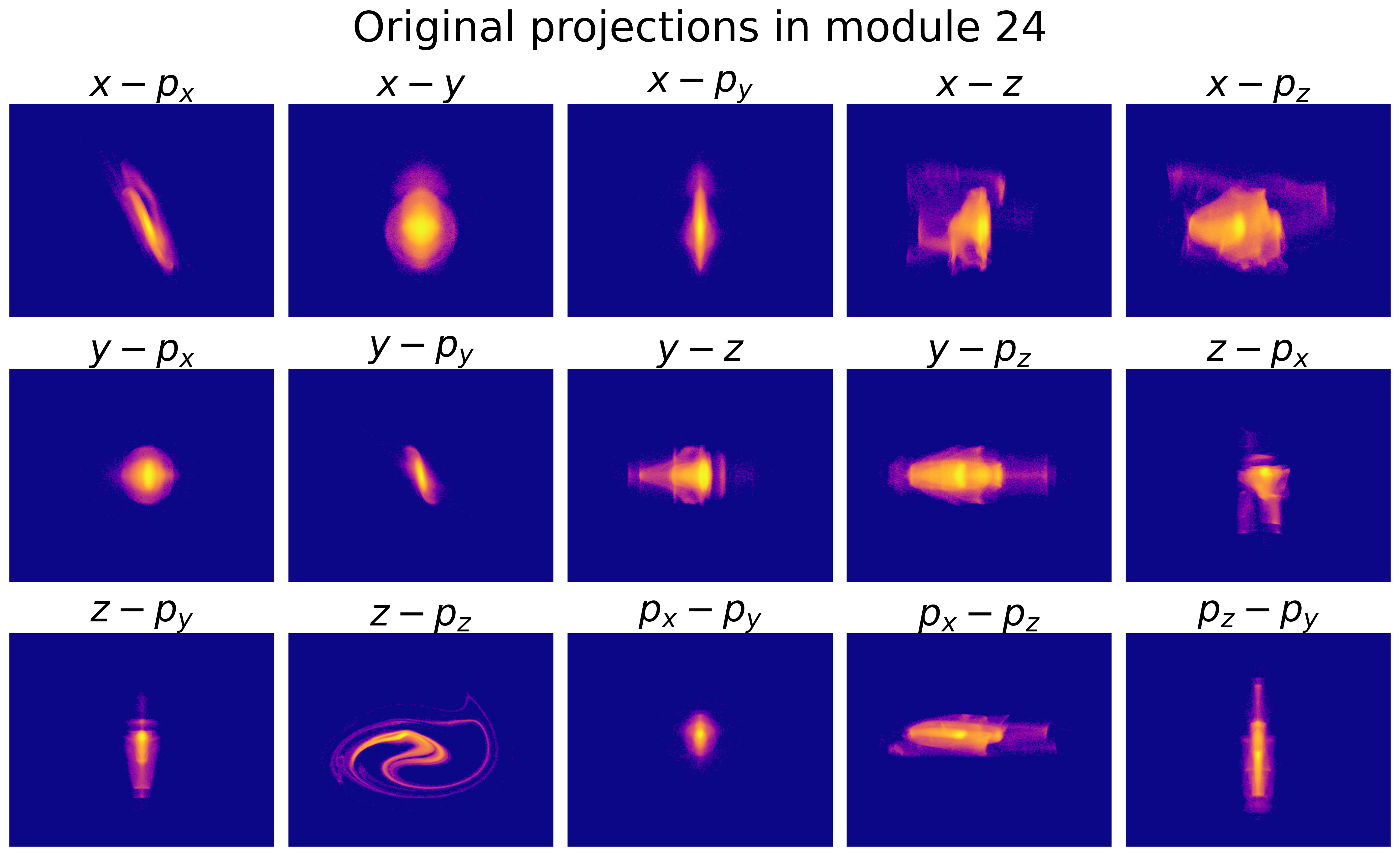}
        \vspace{1mm}
    \end{minipage}
    \begin{minipage}[b]{0.48\linewidth}
        \centering
        \includegraphics[width=1.0\textwidth]{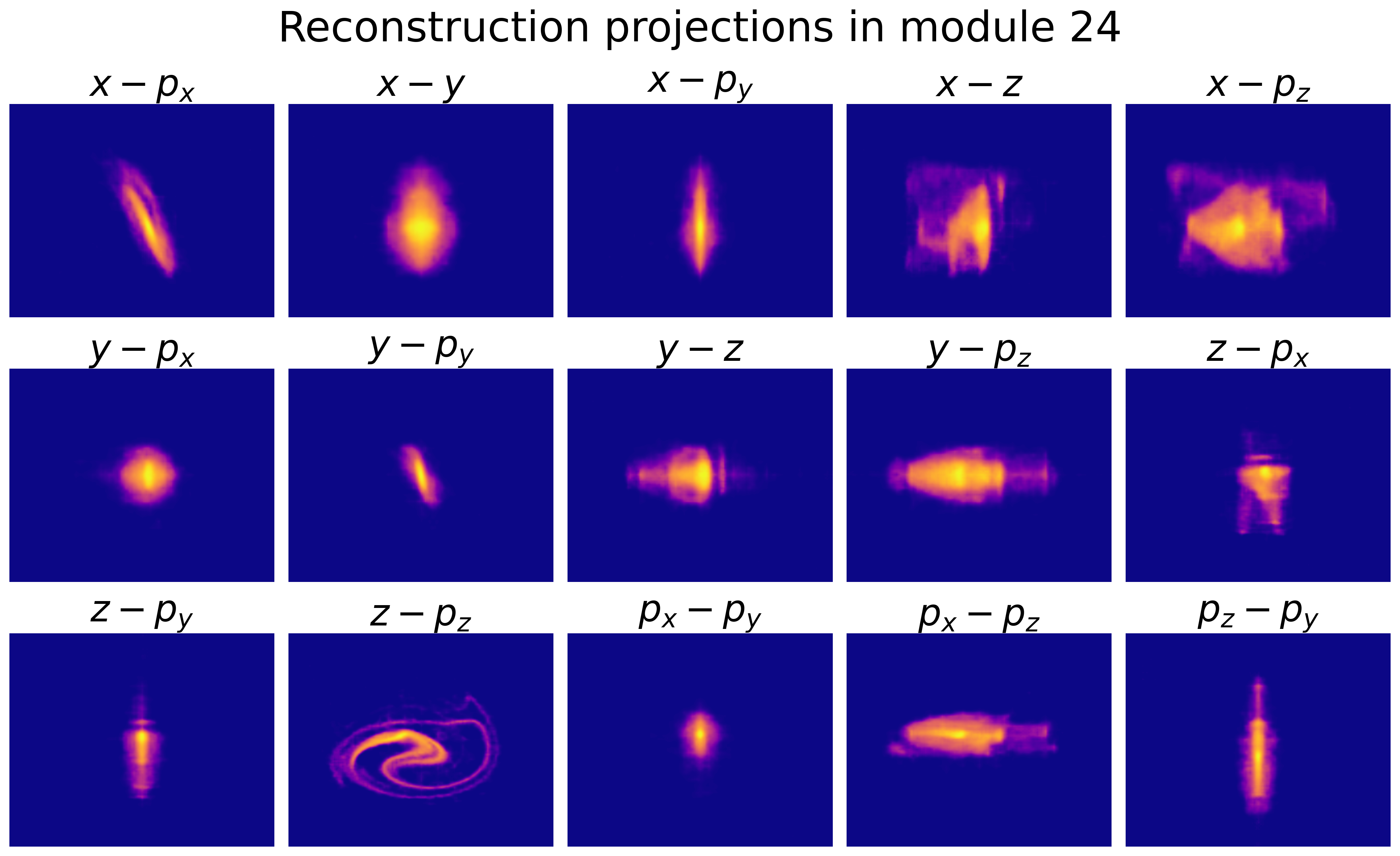}
        \vspace{1mm}
    \end{minipage}
    \begin{minipage}[b]{0.48\linewidth}
        \centering
        \includegraphics[width=1.0\textwidth]{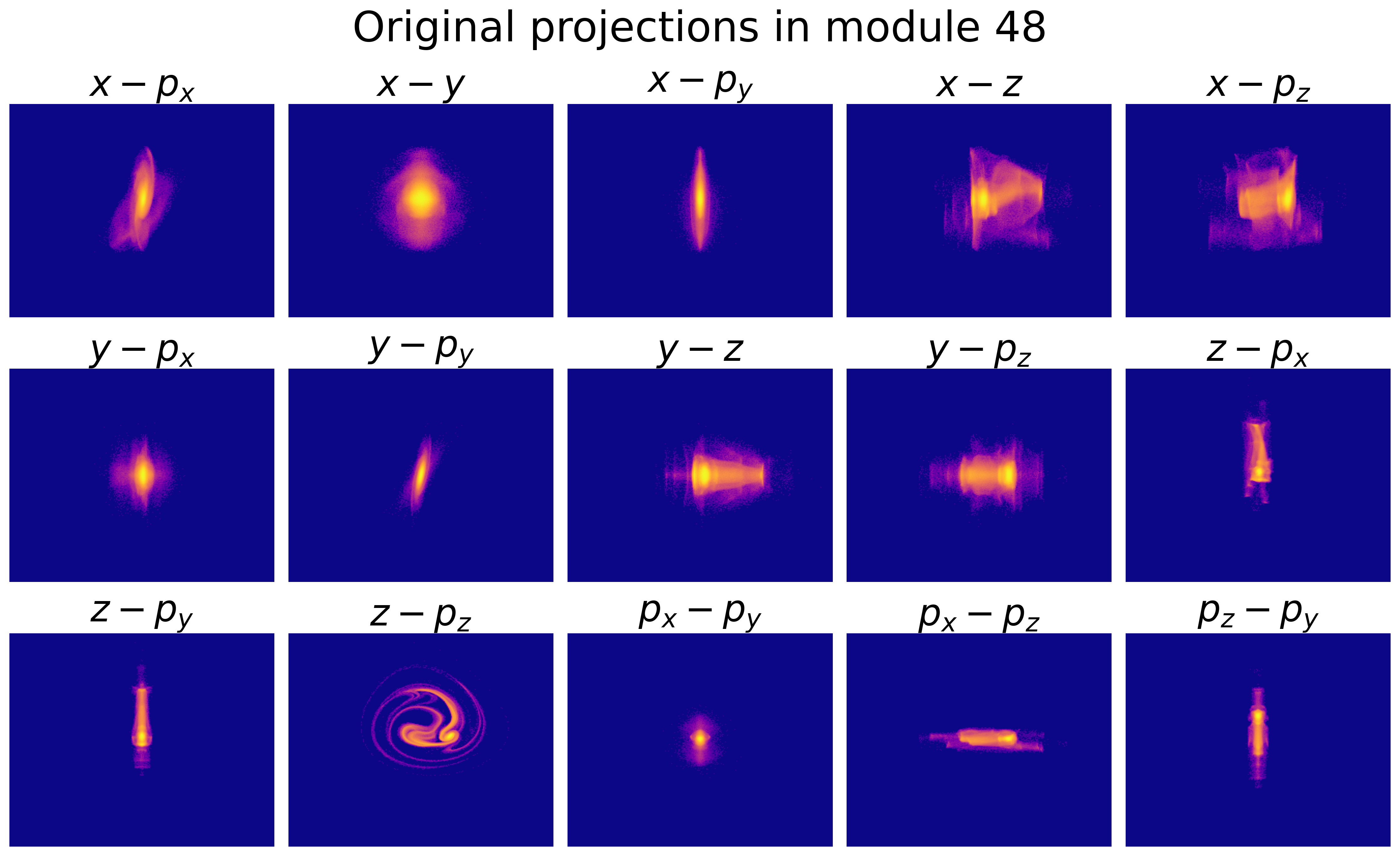}
        \vspace{1mm}
    \end{minipage}
    \begin{minipage}[b]{0.48\linewidth}
        \centering
        \includegraphics[width=1.0\textwidth]{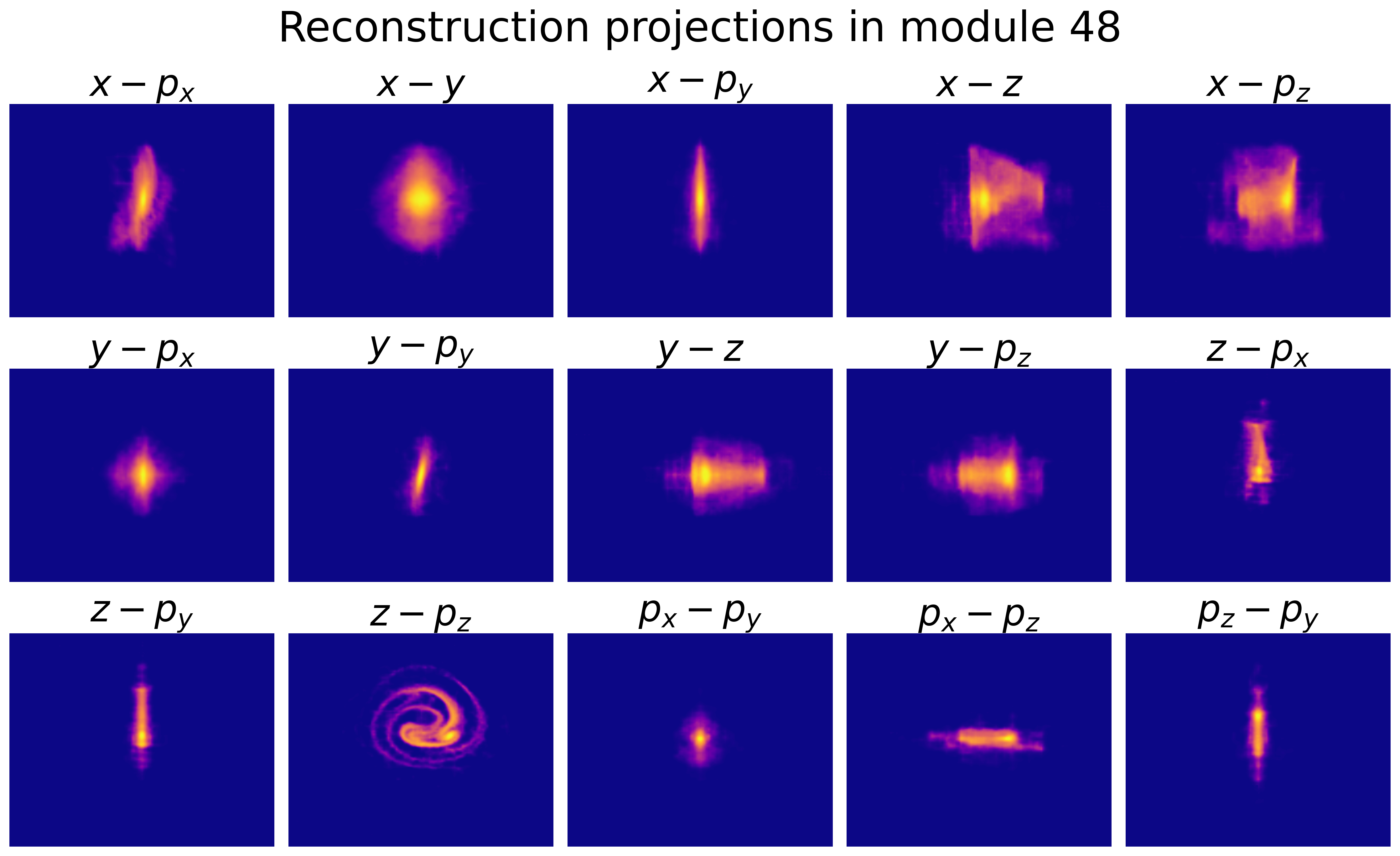}
    \vspace{1mm}
    \end{minipage}
    \begin{minipage}[b]{1.0\linewidth}
        \centering
        \includegraphics[width=1.0\textwidth]{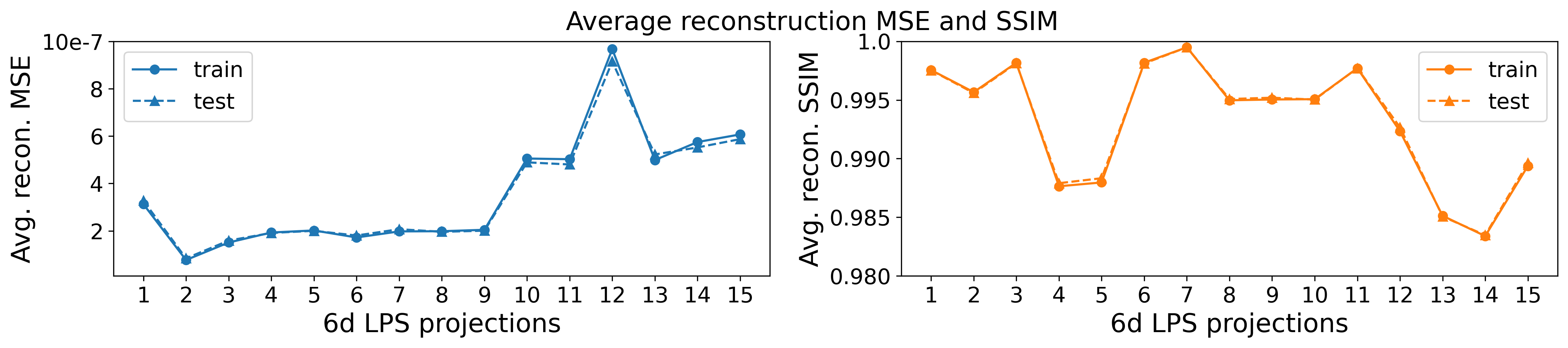}
    \end{minipage}
    \caption{Conditional reconstruction ability of CLARM across three different modules: Left columns (3x5) shows original projections ($X$) from the test dataset whereas right set of columns (3x5) shows the reconstructions ($\hat{X}$) using CVAE. Bottom most plots show average reconstruction MSE and SSIM for different projections across 48 different modules and entire training and test set.}
    \label{fig:reconability}
\end{figure}

Multi-particle tracking simulations are conducted on HPSim to simulate charged particle beam dynamics for the LANSCE linear accelerator. The RF field set points (amplitude and phase) for the DTL (first four modules) are randomly sampled from a 16-dimensional uniform distribution keeping values for RF field set points constant for the remaining accelerating sections. A 96-dimensional vector (the phase and amplitude set points of each of the 48 RF modules) is used as an input to the simulator which outputs a six-dimensional phase space in the form of 15 unique projections at 48 different modules. Each projection is converted into an image (256 $\times$ 256 for this study). We generated 1500 samples of which 1400 are used for training and validation, and 100 are set aside for testing. Each sample contains 48 objects of size 15$\times$256$\times$256 representing all 15 phase space projections at each of the 48 modules. The entire data set has size $1500\times48\times15\times256\times256$. The conditional input $m$ to the encoder is the module number, a scalar between 1-48, normalized to the range $[0,1]$. 

\subsection{Reconstruction ability}
Initially, the first 1400 data objects are divided into training and validation sets with a 85:15 ratio. The CVAE has an encoder with five convolutional layers with 32, 64, 128, 256, 512 filters of size 3$\times$3 each with strides of 2. This is followed by a dense layer with 256 neurons. Additionally, a separate dense branch layer processes the module number with two layers of 32 neurons. Both the primary and branch layers are concatenated and transformed into a 8-dimensional latent space. The decoder mirrors the encoder's architecture with a dense layer followed by 512, 256, 128, 64, and 32 3$\times$3 filters. The activation function is set to LeakyReLU, followed by batch normalization for every layer. The learning rate of 0.001 and a batch size of 32 is found to perform well. The CVAE network is trained with the Adam optimizer for 1500 epochs. As compared to other generative models, CVAE is lightweight model and easier to train.

The reconstruction capability of the trained CVAE is illustrated in Fig.~\ref{fig:reconability} for three different modules (see supplementary Figs. 2-9 for other modules). Fig.~\ref{fig:reconability} shows a comparison between the original and reconstructed phase space projections for the test set. A visual inspection indicates that the trained CVAE effectively reconstructs the projections for the unseen set of inputs. To quantify the quality of reconstruction, we utilized mean squared error (MSE) and structural similarity index (SSIM). MSE captures average local changes in pixels, while SSIM considers luminance, contrast (intensity difference), and structure (correlation) \cite{wang2004image}. SSIM is zero when images are dissimilar, 1 when they are the same, and -1 when the difference is severe. The bottom plots in Fig.~\ref{fig:reconability} depict the average reconstruction MSE and SSIM for 15 different projections across the entire training and test set. The MSE and SSIM values for all projections are promising, confirming the performance of the CVAE in terms of reconstruction ability. It is also observed that MSE values are higher and SSIM values are lower for later projections compared to earlier projections. This is attributed to the contribution from module 1, where the later projections (especially $p_x-p_y$, $p_x-p_z$, $p_z-p_y$) are larger in size and capture a more extensive part of the image. 

\subsection{Latent space visualization}
\begin{figure}[hbtp!]
    \centering
    \begin{minipage}[b]{1.0\linewidth}
        \centering
        \includegraphics[width=0.9\textwidth]{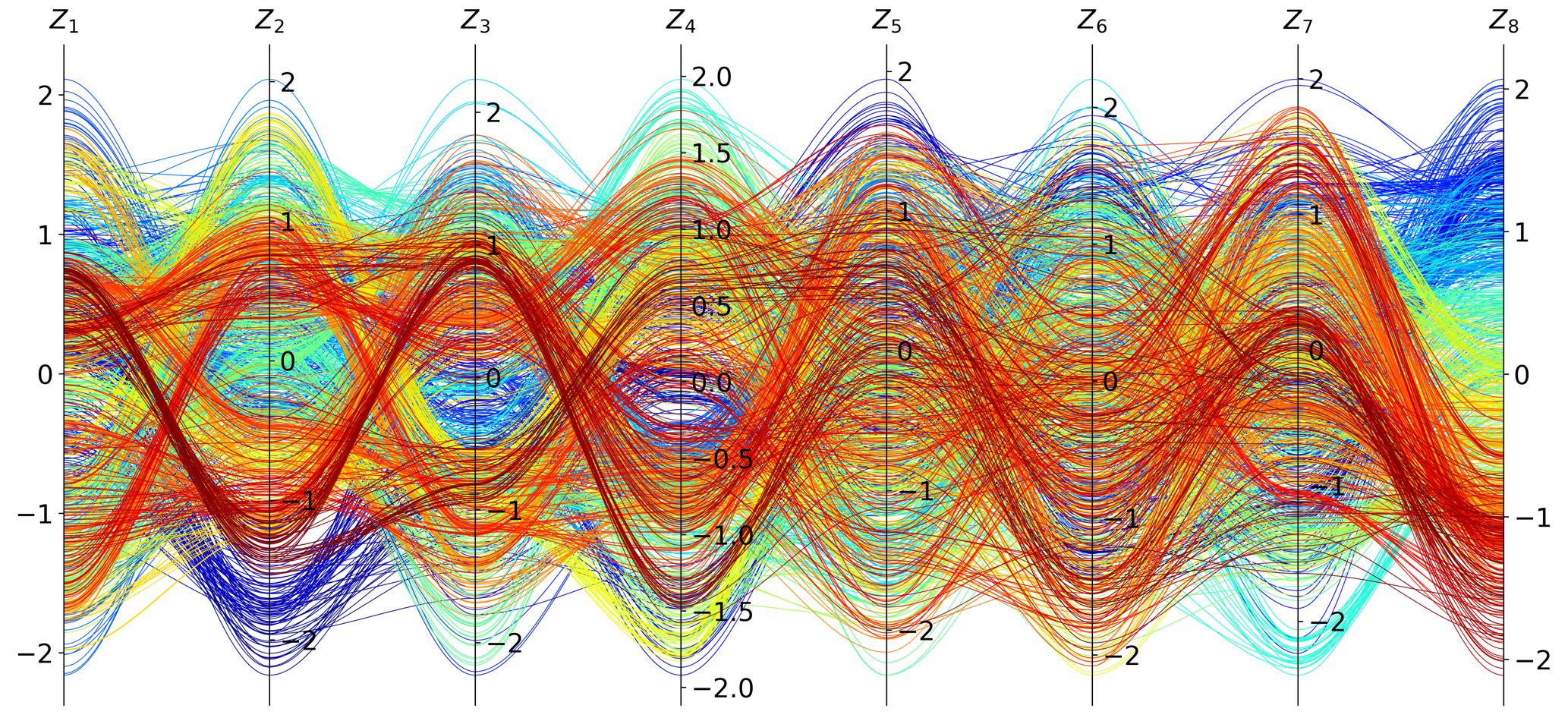}
    \end{minipage}
        \begin{minipage}[b]{0.24\linewidth}
        \centering
        \includegraphics[width=1.0\textwidth]{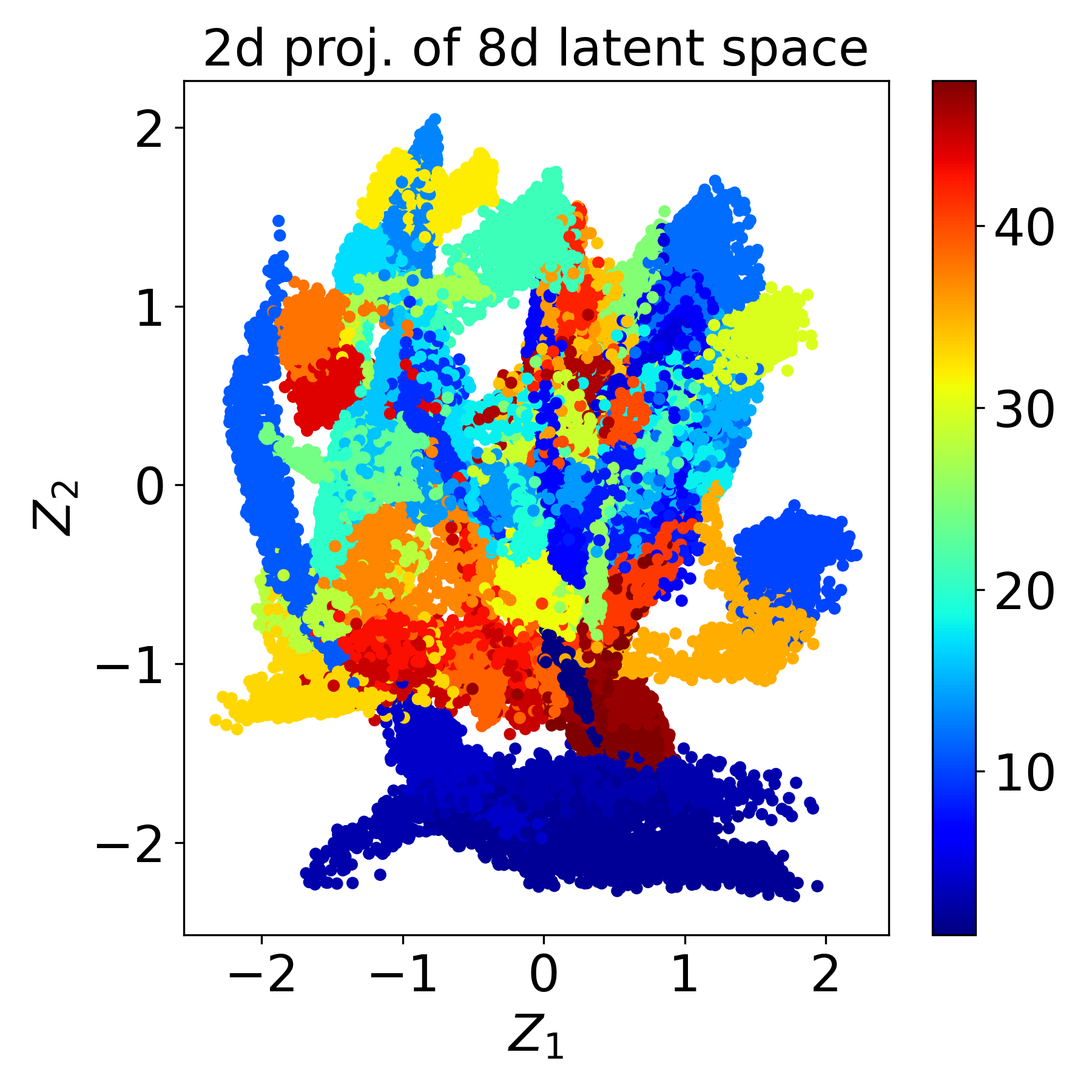}
    \end{minipage}
    \begin{minipage}[b]{0.24\linewidth}
        \centering
        \includegraphics[width=1.0\textwidth]{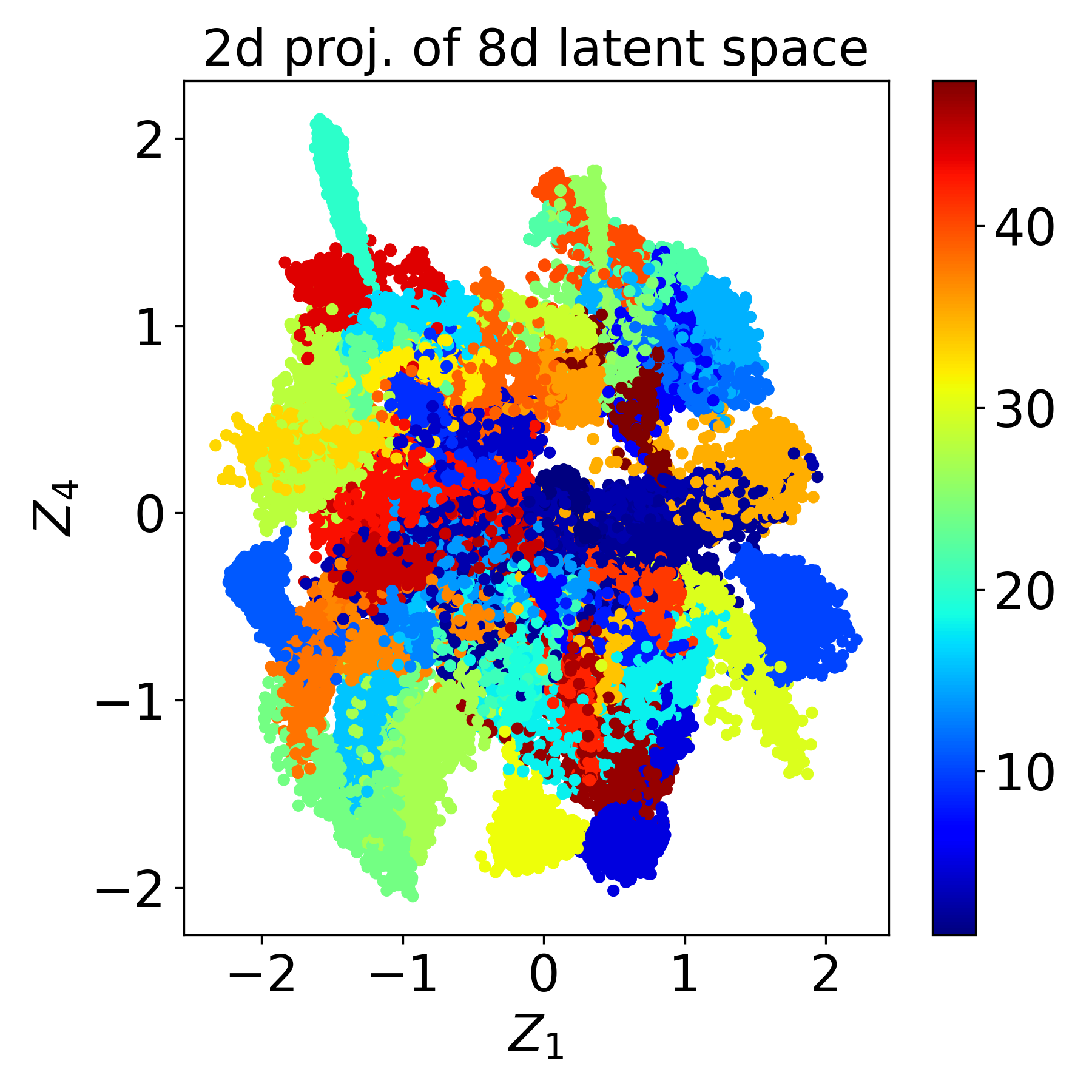}
    \end{minipage}
    \begin{minipage}[b]{0.24\linewidth}
        \centering
        \includegraphics[width=1.0\textwidth]{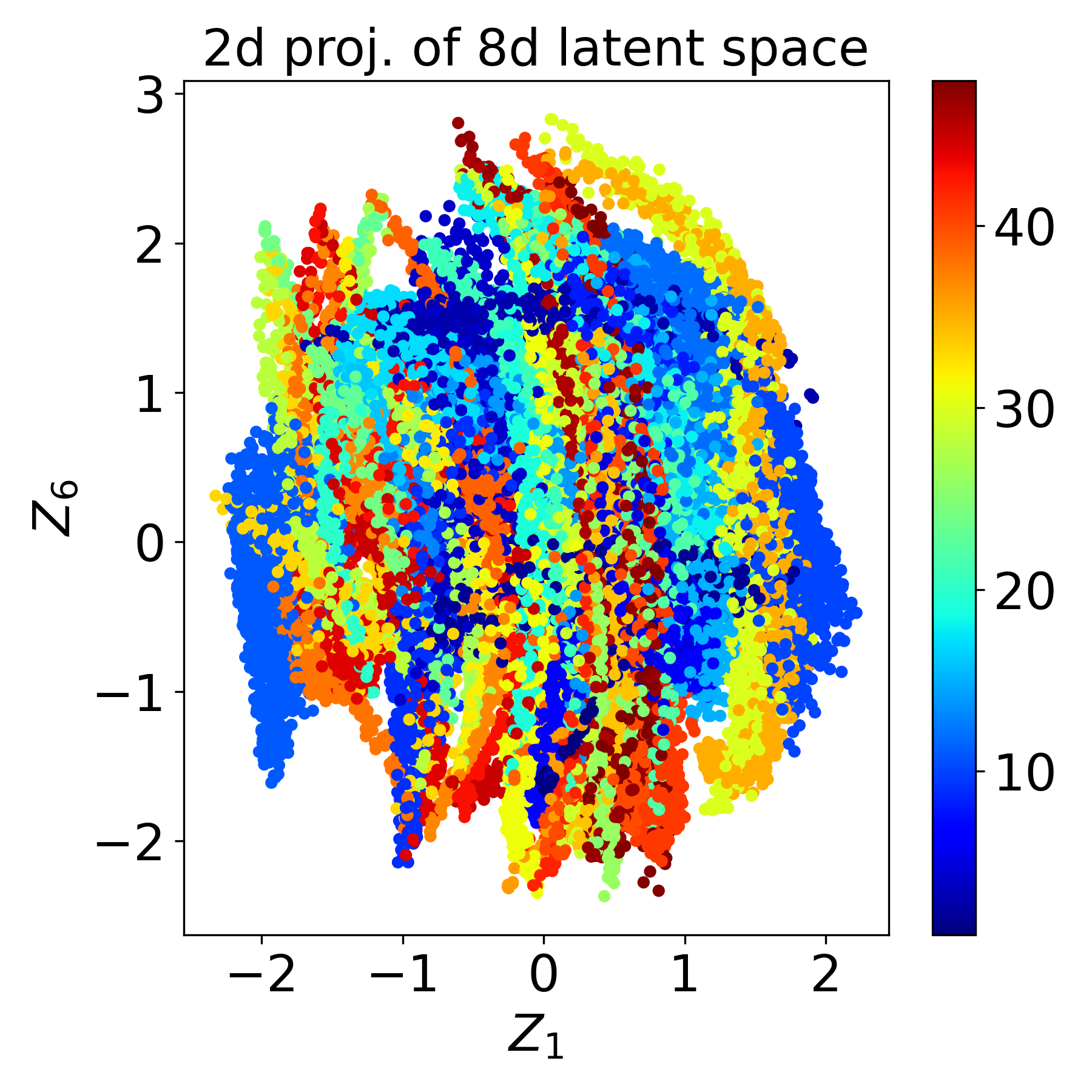}
    \end{minipage}
    \begin{minipage}[b]{0.24\linewidth}
        \centering
        \includegraphics[width=1.0\textwidth]{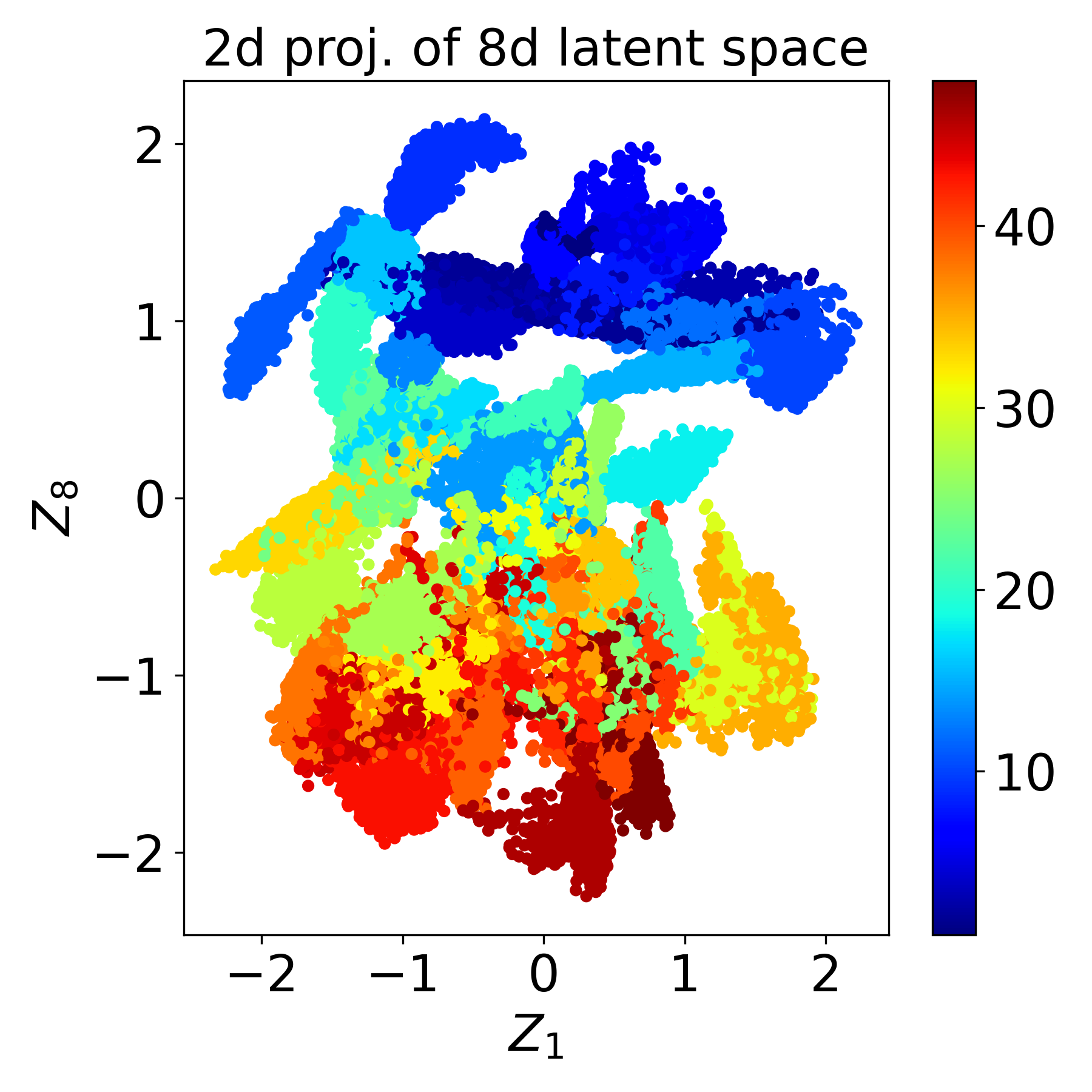}
    \end{minipage}
    \begin{minipage}[b]{0.32\linewidth}
        \centering
        \includegraphics[width=1.0\textwidth]{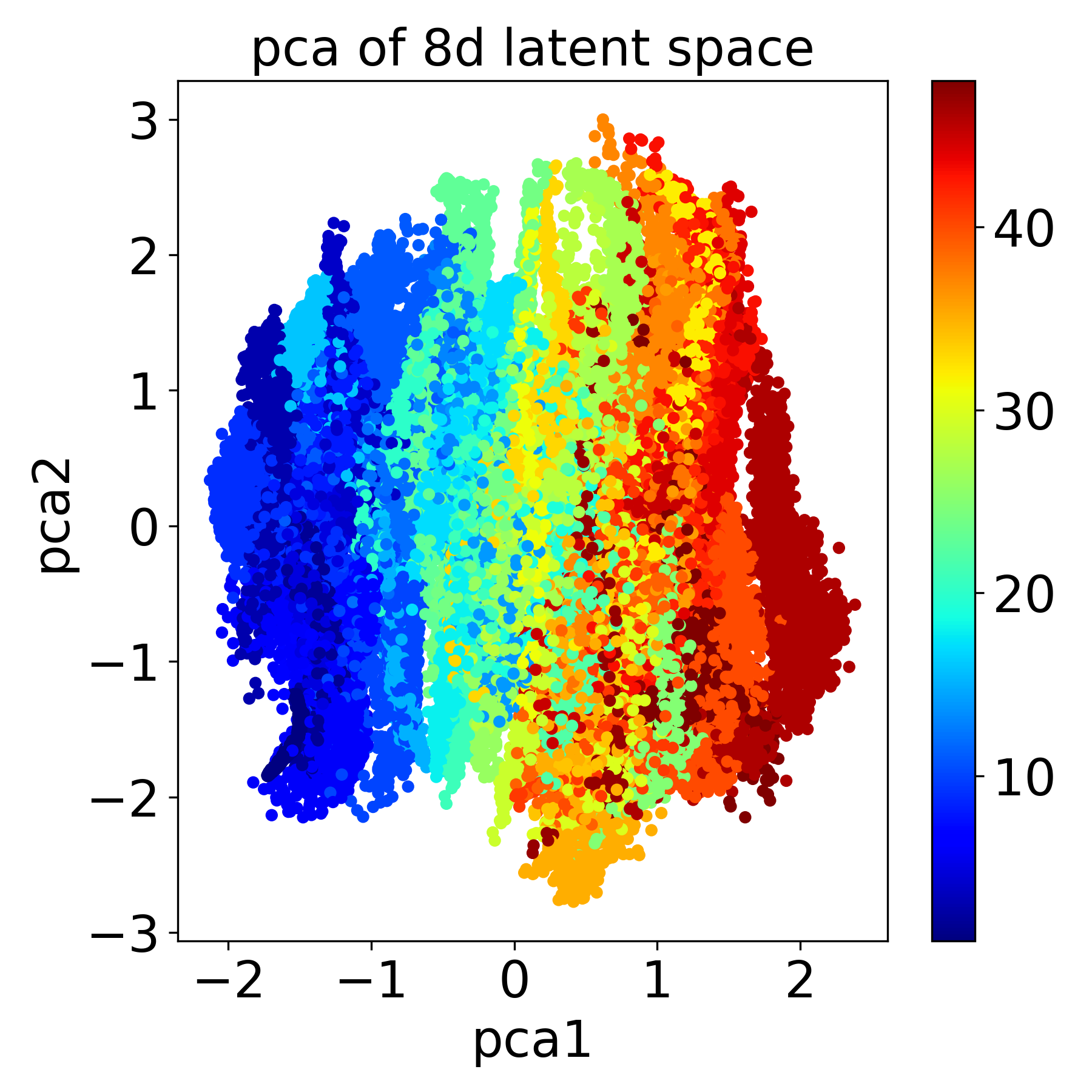}
    \end{minipage}
    \begin{minipage}[b]{0.32\linewidth}
        \centering
        \includegraphics[width=1.0\textwidth]{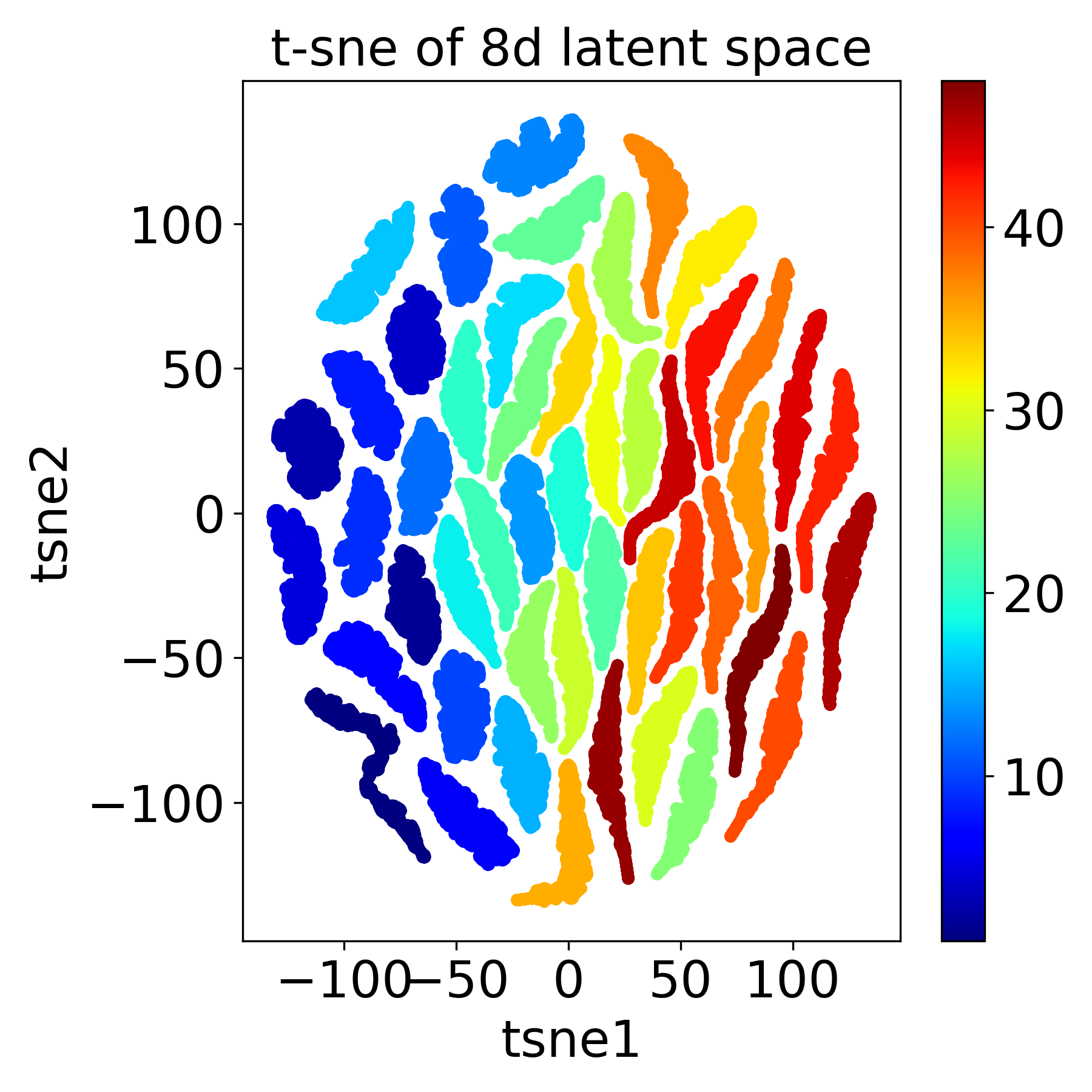}
    \end{minipage}
    \begin{minipage}[b]{0.32\linewidth}
        \centering
        \includegraphics[width=1.0\textwidth]{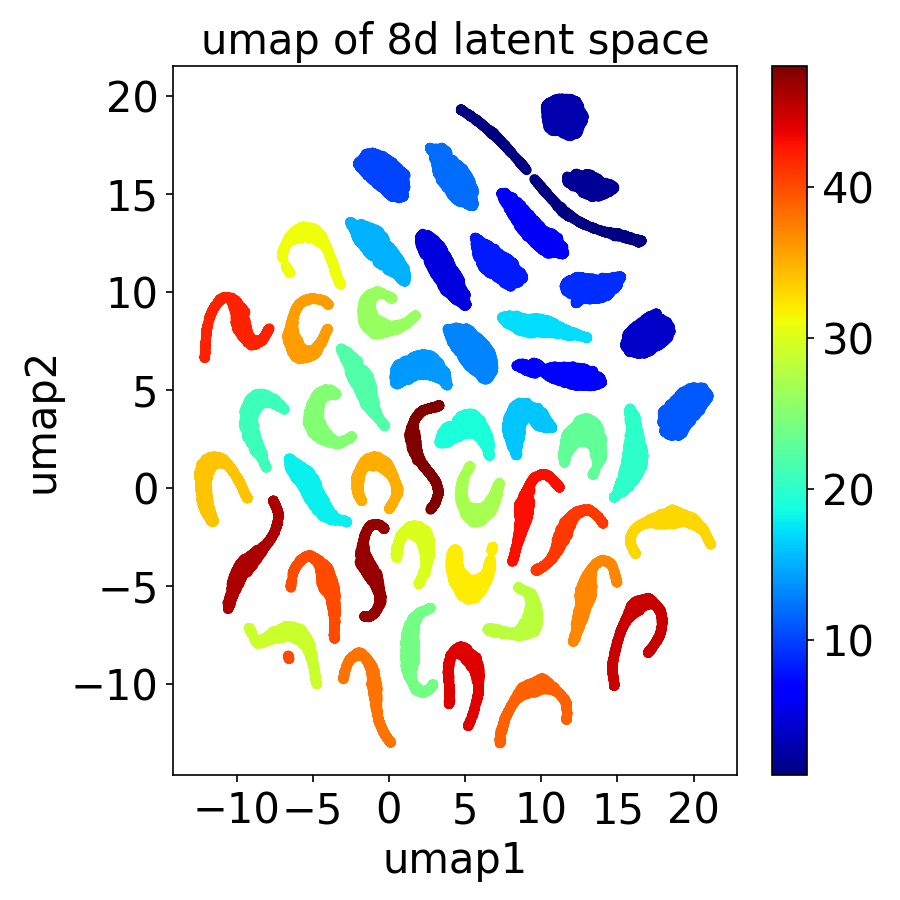}
    \end{minipage}
    \caption{Latent space visualization of CVAE: The top row shows 8D latent space using parallel coordinates plots where different colors corresponds to different modules and every curve is a point in the 8D latent space. The middle row shows various 2D projections of the 8D latent space ($Z_1 - Z_2$, $Z_1 - Z_4$, $Z_1 - Z_6$, $Z_1 - Z_8$). The bottom row shows the 2D PCA, t-SNE and UMAP of 8D latent space. The color maps for all of the rows are the same, corresponding to module number.}
    \label{fig:latentvisualization}
\end{figure}

Visualization of the latent space provides insights into the extracted features and improves the interpretability of the network. The 8D latent space is plotted using a parallel coordinates plot in the top row of Fig.~\ref{fig:latentvisualization}. To make the plot more clear, we have plotted only 2500 randomly selected points from the entire dataset. Each curve in the plot represents a point in 8D space, and the different colors of the curves correspond to different modules (refer to the colorbar). The middle row of Fig.~\ref{fig:latentvisualization} shows four different 2D projections of the 8D latent space. It can be seen that the projections in modules are much more separated in $Z_1-Z_8$ than $Z_1-Z_2$, and $Z_1-Z_4$  and $Z_1-Z_5$ projections. Similar separation can also be seen in the top row of Fig.~\ref{fig:latentvisualization} along $Z_8$. Such visual analysis can also be extended to the other 24 projections, however, the visualization is still restricted by the high dimensionality of the latent space. To further enhance the visualization, the 8D latent space is transformed into various different 2D spaces, as shown in the bottom row of Fig.~\ref{fig:latentvisualization}. The first method is a linear dimensionality reduction technique called principal component analysis (PCA). The other two methods are manifold learning techniques called t-distributed Stochastic Neighbor Embedding (t-SNE) \cite{van2008visualizing} and Uniform Manifold Approximation and Projection (UMAP) \cite{mcinnes2018umap}. Both of them are non-linear dimensionality reduction approaches, contrary to PCA. While t-SNE is a more popular method for various problems, UMAP performs better in preserving both local and global structures, computational efficiency, and parameter robustness \cite{mcinnes2018umap}. 

It is seen from the PCA plot that the points belonging to the initial and end modules are well separate, occupying the left and right ends of the plot. However, the middle modules are coarsely separated with a lot of intermixing of points of different modules. This is to be expected as PCA is a linear method that is unable to separate this 8D data set when using only 2 principal components. When fitting the PCA projection matrix, it was found that no $n<8$ dimensional PCA representation resulted in explained variance $>95\%$ , which gives us confidence that the encoder is fully utilizing the entire 8D latent space. This implies that increasing the dimensionality of the latent space could make it easier for the encoder to compress and for the decoder to then reconstruct the high resolution stacks of 15 images. This brings up the choice of dimensionality of the latent space. There is a tradeoff that has to be considered, as the latent space dimension becomes smaller, training the encoder and the decoder becomes more challenging, but it makes the LSTM network's job easier as it is learning the temporal dynamics of a lower dimensional system. By trying various latent space dimensions (starting at $\mathbf{z}\in\mathbb{R}^2$), we found that 8D was a good tradeoff between representation power and training speed and accuracy. 

As expected, the nonlinear t-SNE and UMAP methods show better separation between different modules. In t-SNE, similar to PCA, the initial and end modules occupy different ends of the plot, which is not true for UMAP. However, the UMAP shows better separation among the modules than t-SNE. It is easily noticeable that t-SNE and UMAP perform better than PCA. This is because PCA is formulated to transform the high-dimensional data into a lower-dimensional space while retaining as much information as possible in the form of maximizing the variance of the data. On the other hand, t-SNE and UMAP are designed for better clustering of points in the lower dimensional space.

\subsection{Generative ability}
One of the nice features of the CVAE method is a well behaved latent space without separate disjoint regions, which can be continuously traversed to provide accurate samples, represented as a learned lower-dimensional distribution or density explicitly. This distribution can be randomly sampled and processed through the trained decoder, enabling the generation of new realistic phase space projections across different modules. In comparison to Generative Adversarial Networks (GANs), VAEs, including CVAE, avoid issues like mode collapse and are generally easier to model and train more efficiently \cite{rautela2022towards}. The ability to generate new data is crucial not only for increasing the dataset size for robust diagnostics but also for problems like system parameter estimation and control \cite{scheinker2023adaptive,williams2023experimental}. In our approach, we have utilized the latent space of CVAE to generate new, realistic phase space projections across different modules. The latent space is randomly sampled within the bounds of a specific module, and the sampled point is then processed through the decoder to generate new projections. 

\begin{figure}[htbp!]
    \centering
    \begin{minipage}[b]{1.0\linewidth}
        \centering
        \includegraphics[width=1.0\textwidth]{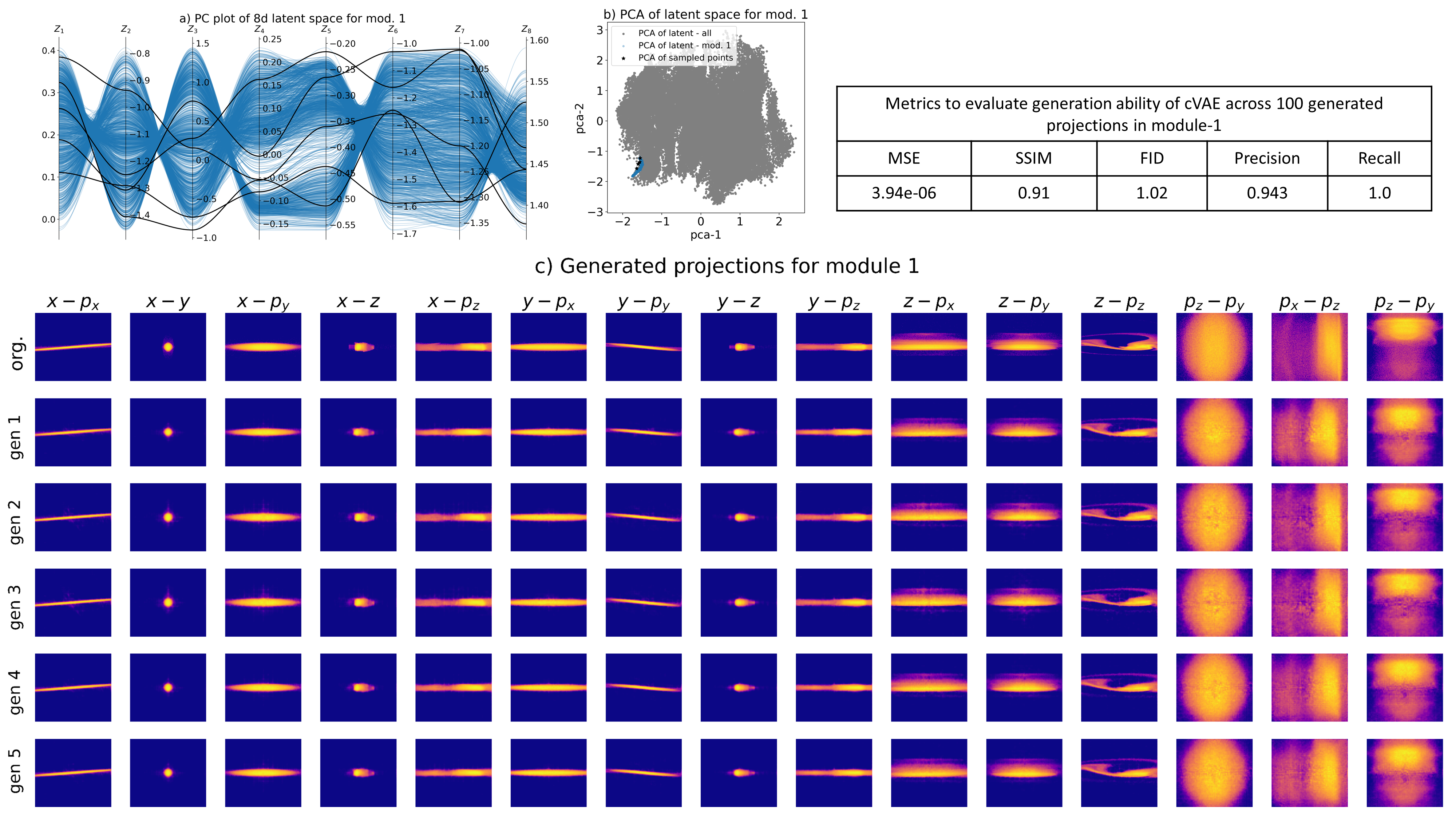}
    \end{minipage}
    \begin{minipage}[b]{1.0\linewidth}
        \centering
        \includegraphics[width=1.0\textwidth]{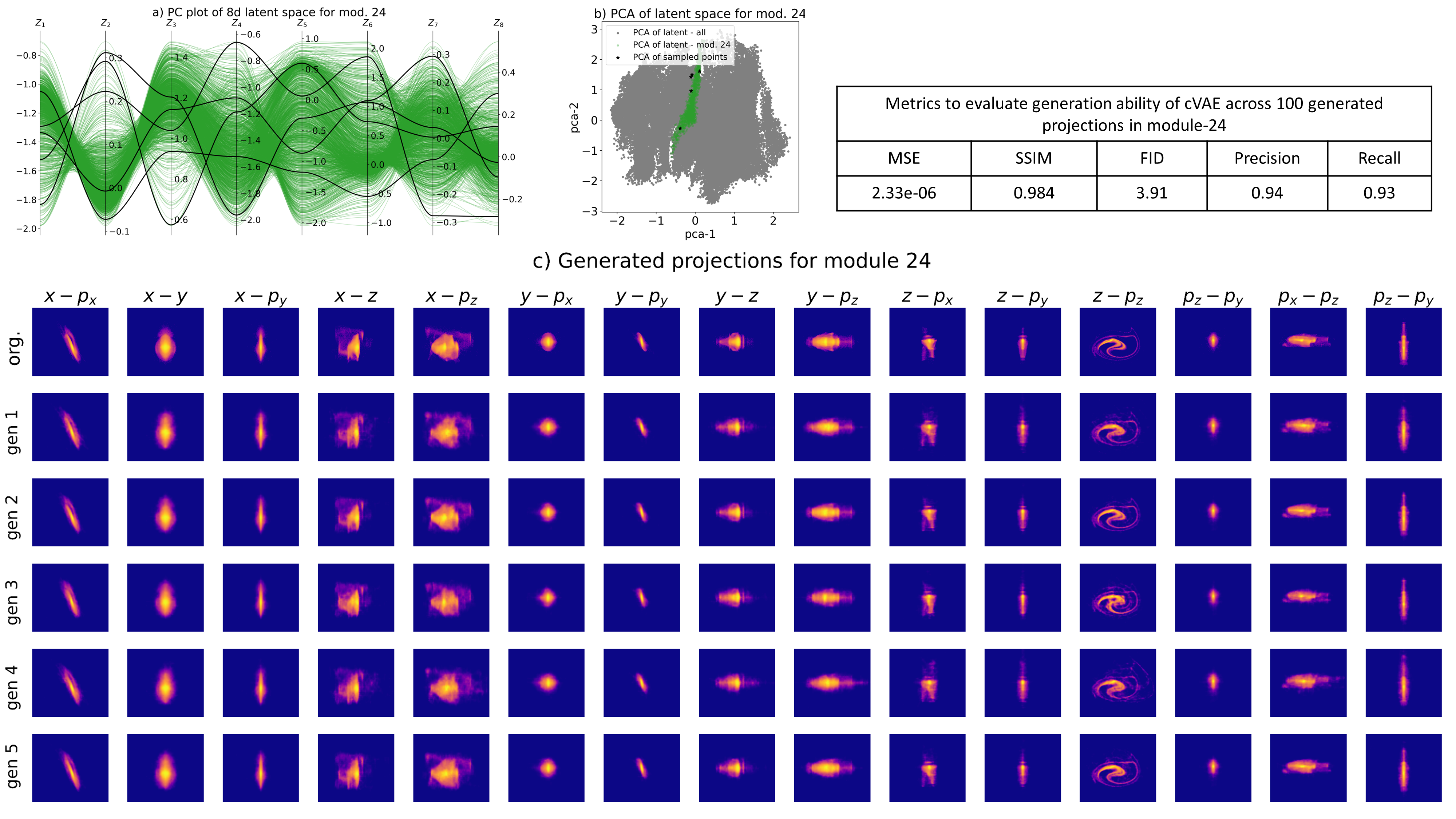}
    \end{minipage}
    \begin{minipage}[b]{1.0\linewidth}
        \centering
        \includegraphics[width=1.0\textwidth]{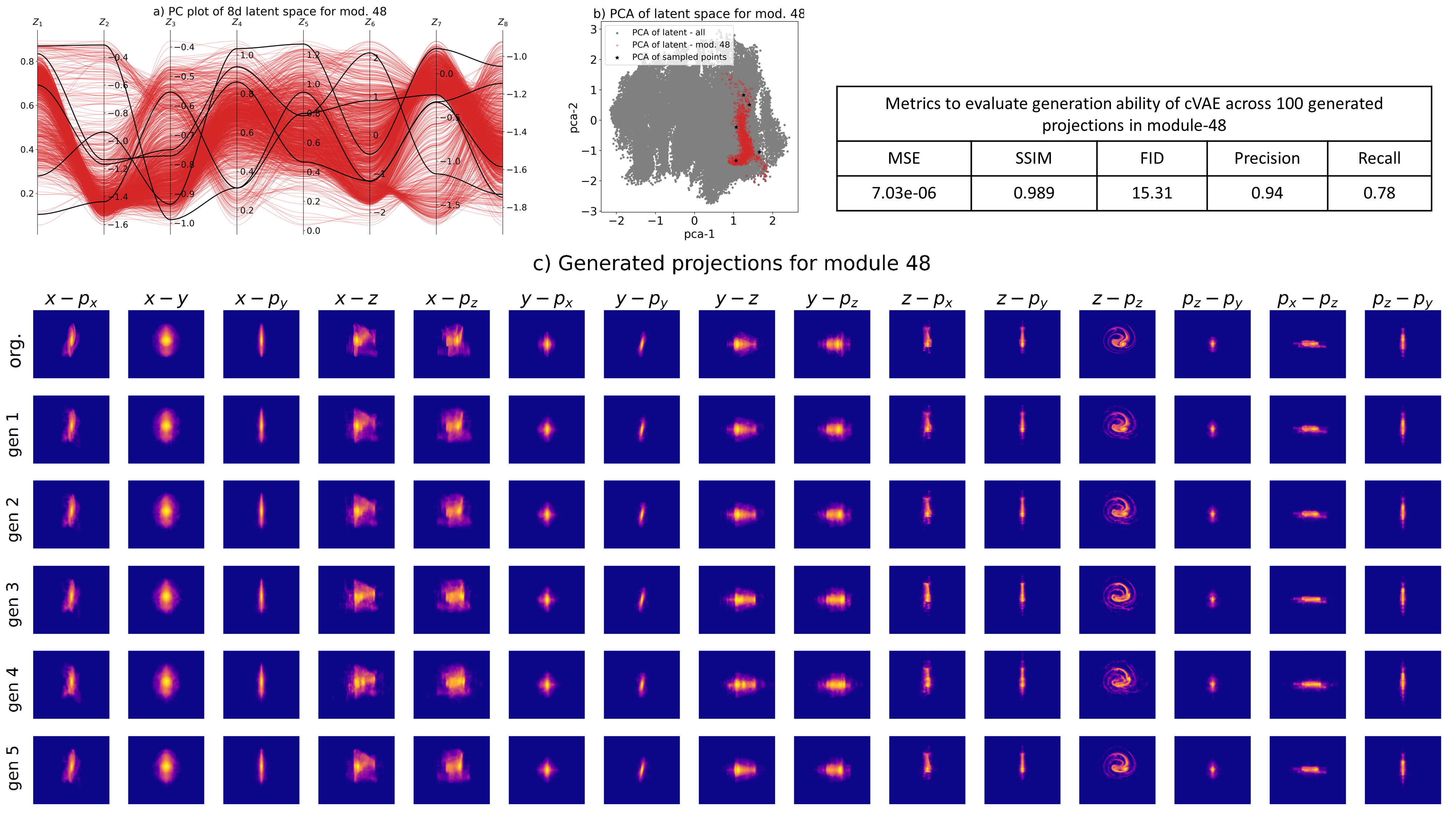}
    \end{minipage}
    \caption{Conditional generation ability across different modules (1,24,48) by conditional sampling the latent space randomly and using decoder to generate the projections: (a) PC plot showing 5 sampled points in black curves and all the points of the corresponding module in blue, green and red, (b) PCA of 5 8-dimensional latent space in blue, green and red and sampled points in black, (c) - original projections from test set and 5 generated projections. The tabular form showing metrics to evaluated the generative ability: MSE, SSIM, FID, Precision and recall.}
    \label{fig:generativeability}
\end{figure}

Fig.~\ref{fig:generativeability}c presents five generated phase space projections for module numbers 1, 24, and 48 (see supplementary Figs. 10-17 for other modules). These generated projections correspond to five randomly sampled points (highlighted in black color) of the latent space, shown in both the parallel coordinates (Fig.~\ref{fig:generativeability}a)) and PCA (Fig.~\ref{fig:generativeability}b)) plots. The background color (blue, green, red) of the curves (in the PC plot) and points (in the PCA plot) represents the entire latent subspace of the module. It is observed that the generated projections look realistic and promising upon visual inspection. To quantify the generative ability of the CVAE, we have used five different evaluation metrics predominately used for generative models. In addition to the MSE and SSIM employed previously, we have introduced the Frechet Inception Distance (FID), precision, and recall for generation evaluation. The FID metric is based on Frechet distance to measure the distance between two probability distributions (real and generated) in a metric space \cite{heusel2017gans}. Mathematically, $FID = \vert \mu_1-\mu_2 \vert + Tr(\sigma_1 + \sigma_2 - 2\sqrt(\sigma_1\ast\sigma_2))$, where $\mu_1$, $\sigma_1$ and $\mu_2$, $\sigma_2$ are the mean and standard deviation of real and generated distributions and $Tr$ is the trace.  Lower FID values correspond to a higher quality of the generated images. Precision and recall, although primarily used for classification, can be adapted to evaluate the distribution of generated projections \cite{sajjadi2018assessing}. Precision represents the ratio of correctly generated projections to the total number of generated projections, while recall represents the ratio of correctly generated projections to the total number of real samples. Higher precision indicates that the distribution of generative projections closely resembles the distribution of real projections. On the other hand, high recall signifies that the generated samples span the diversity of the real distribution \cite{giannone2023aligning}. 

In the literature, various models such as InceptionV3 and Resnet50 are commonly employed to compute FID, precision, and recall. Calculating FID relies on the features extracted from a trained model (at a particular layer) corresponding to real and generated images. However, it is important to note that these models, including InceptionV3 and Resnet50, are typically trained on the ImageNet dataset, which comprises 1.2 million RGB images across 1000 classes of real-world objects. In contrast, our dataset is very different, focusing on capturing charged particle dynamics in linear accelerators. Our dataset contains images with 15 channels associated with 48 classes. For our study, we adapted the architecture of the Resnet50 model, modifying it to accommodate a 15-channel input and provide a 48-class output \cite{rezende2017malicious}. This modified Resnet50 classifier was trained from scratch on our dataset to classify projections into 48 classes, corresponding to various accelerating sections/modules. The classifier is trained for 200 epochs, achieving outstanding accuracy with 1.0 on the training and validation sets and 0.9998 on the test set.

The trained Resnet50 model is then utilized to calculate FID, precision, and recall metrics. Specifically, we extracted 2048 features for both real and generated images corresponding to the final average pooling layer just before the flattening layer \cite{he2016deep}. These features are then utilized to calculate the FID score for 100 sets of real and generated projections for every module, as detailed in the tabular format in Fig.~\ref{fig:generativeability}. The FID score is regarded as a subjective metric, influenced by the model, dataset, and the quantity of generated samples \cite{cheng2020data}. In our study, we have observed variability in the FID score based on the number of generated samples. It is noted that FID scores stabilize as the number of generated samples approaches a hundred. We have reported a maximum FID score of approximately 15, and it's worth noting that FID scores of 15 or lower are generally considered good for generative models \cite{cheng2020data}. Precision and recall metrics are also computed using the entire model as a classifier for 100 real and generated projections, and the results are presented in the table. Additionally, the table includes the average MSE and SSIM for the generated projections in different modules.  The obtained metrics exhibit promising results, with lower MSE and FID values, and higher SSIM, precision, and recall values, affirming the generative ability of the network. Another interesting observation is that the FID score is higher, while precision and recall scores are lower for later modules compared to initial modules. This is attributed to the lower variations in the projections across the later modules, as discussed in Sec.~\ref{ssec:simulations}. Projections that share a similar appearance result in the intersection of their corresponding subspaces in the latent space. When a random sample from the latent space emanates from a combination of multiple subspaces, the generated projections of a particular module may incorporate features from projections of other modules. This results in a higher FID score, and lower precision and recall.

\subsection{Forecasting ability}
An LSTM network is trained on the latent space of the CVAE to forecast the 8D latent vectors of subsequent modules based on the 8D latent vectors of previous modules used as inputs. Initially, the latent space data is reshaped back from 67,200$\times$8 to 1400$\times$48$\times$8. Here, 1400 signifies the number of samples, 48 denotes the trajectory length (representing the total number of modules), and 8 represents the number of features (dimension of the latent space) for the LSTM. The LSTM network is designed to handle variable-length sets of 8D latent vectors as inputs and a single latent vector is produced as an output. In our case, the input size can vary from $4\times8$, $5\times8$, .., $47\times8$, while the output remains fixed at ${1\times8}$ (representing the next module). This approach enables the network to learn the autoregressive nature of the latent points, crucial for formulating the autoregressive loop of CLARM. The LSTM architecture consists of 2 layers and 64 hidden units in each layer. Training involves minimizing the mean squared loss between true and forecasted latent vectors, with Adam acting as the optimizer. A learning rate of 0.001 is chosen, and a batch size of 16 is used for training.

\begin{figure}[htbp!]
\centering
\includegraphics[width=1.0\linewidth]{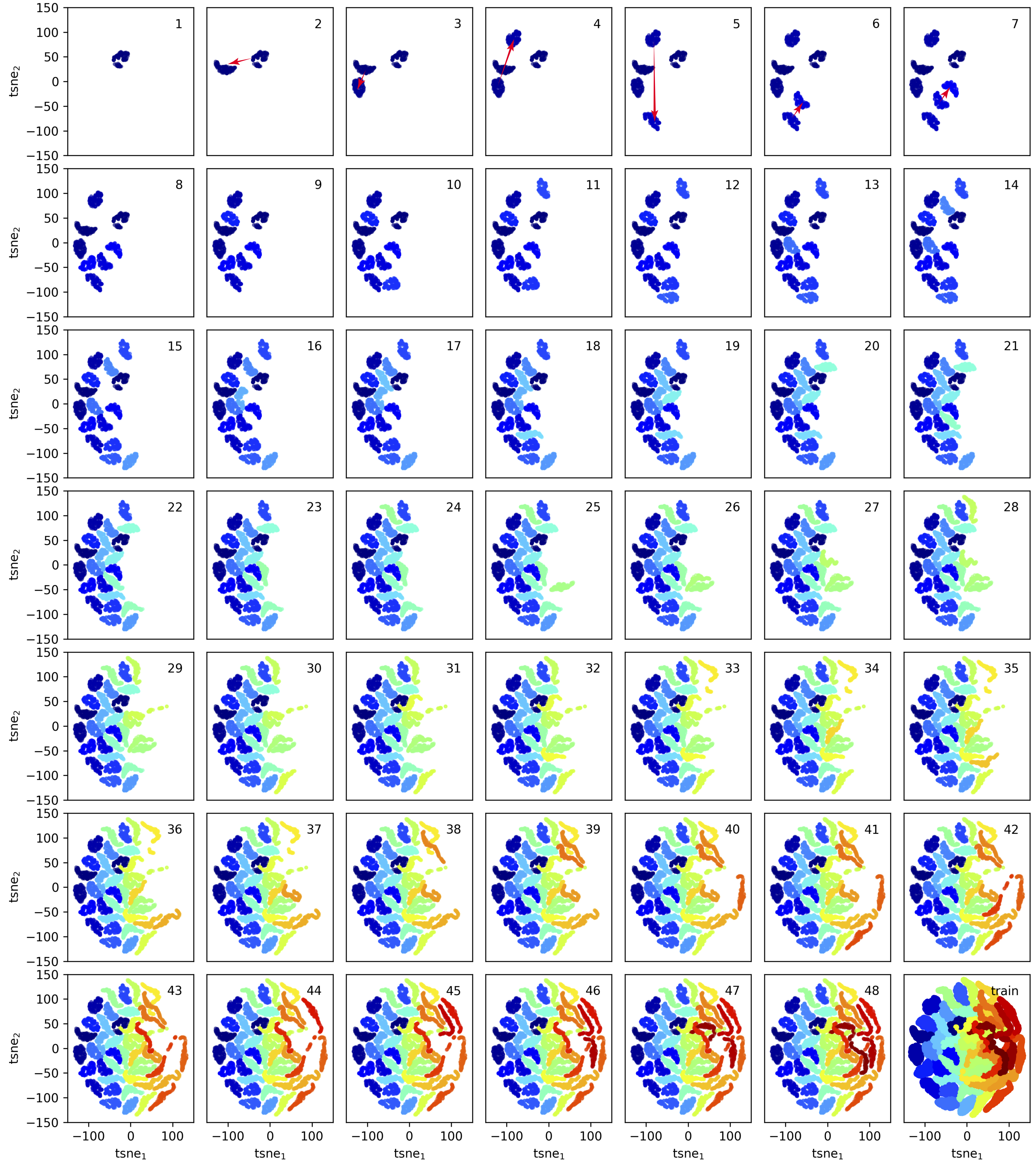}
\caption{Latent space trajectories: t-SNE based visualization of the LSTM-based forecasting of phase space projections in the latent space for the entire test data set (of size 100). Plot 1 shows the location where the encoder maps all 100 test data points of Module 1. Plot 2 shows where those 100 points move to when passing to Module 2, with a red arrow. Plots 3-7 show subsequent movement up to Module 7. The remaining plots show the same procedure until the locations of all 48 modules for all of the test data is built up (arrows removed starting with moudle 8 or else the figure gets too messy). Plot 48 provides the fully evolved latent space at the end of the LSTM-based forecasting algorithm. Plot 49 shows the t-SNE of the latent space for the entire training data set. The t-SNE of the fully evolved latent space for the given test set is structurally very similar to the t-SNE of the latent space of the training set.}
\label{fig:latenttraj}
\end{figure}

Algorithm~\ref{alg:predict} and Sec.~\ref{ssec:theory} describes the forecasting methodology. In Fig.~\ref{fig:latenttraj}, we employ t-SNE for the visualization of latent space evolution while performing forecasting. The figure captures all the forecasted latent points in the current and previous modules. It can be observed that the fully evolved latent space in the plot exhibits a close resemblance to the latent space of the training set.

\begin{figure}[htbp!]
    \centering
    \begin{minipage}[b]{0.48\linewidth}
        \centering
        \includegraphics[width=1.0\textwidth]{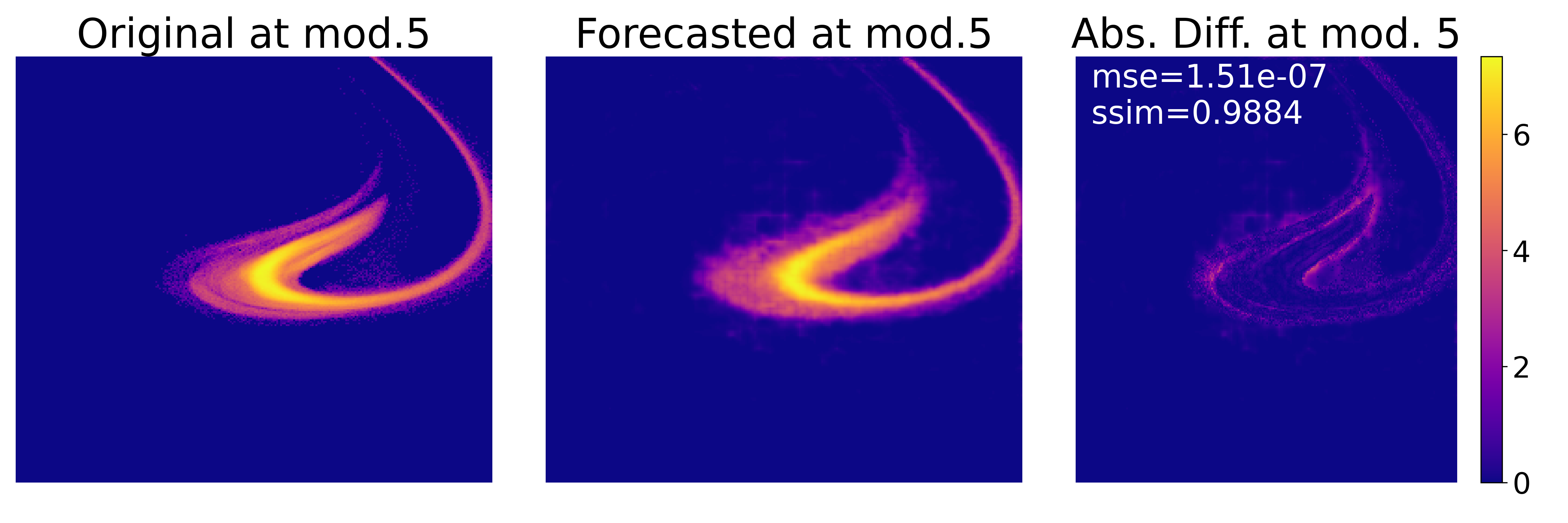}
    \end{minipage}
    \begin{minipage}[b]{0.48\linewidth}
        \centering
        \includegraphics[width=1.0\textwidth]{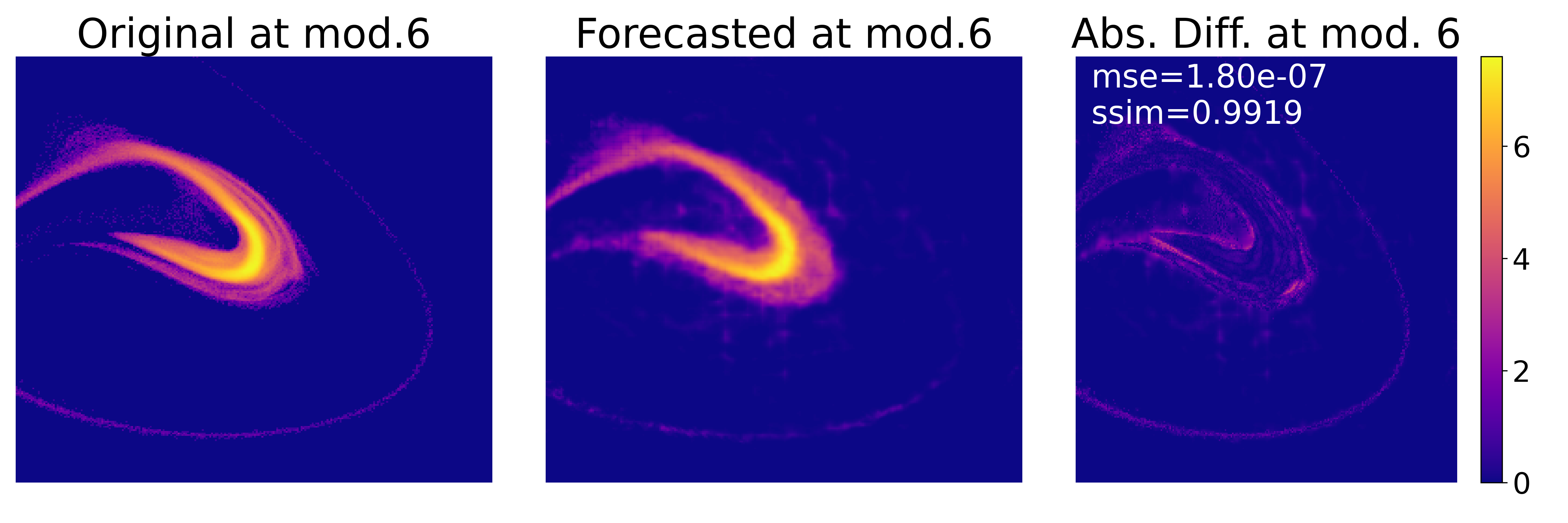}
    \end{minipage}
    \begin{minipage}[b]{0.48\linewidth}
        \centering
        \includegraphics[width=1.0\textwidth]{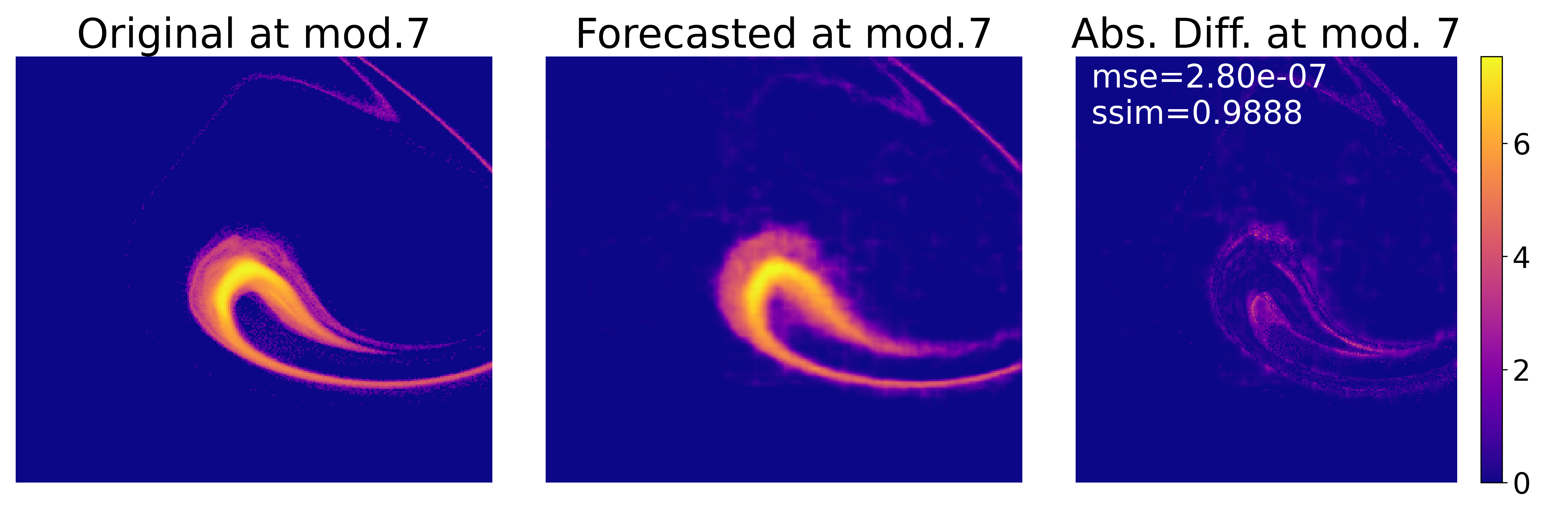}
    \end{minipage}
    \begin{minipage}[b]{0.48\linewidth}
        \centering
        \includegraphics[width=1.0\textwidth]{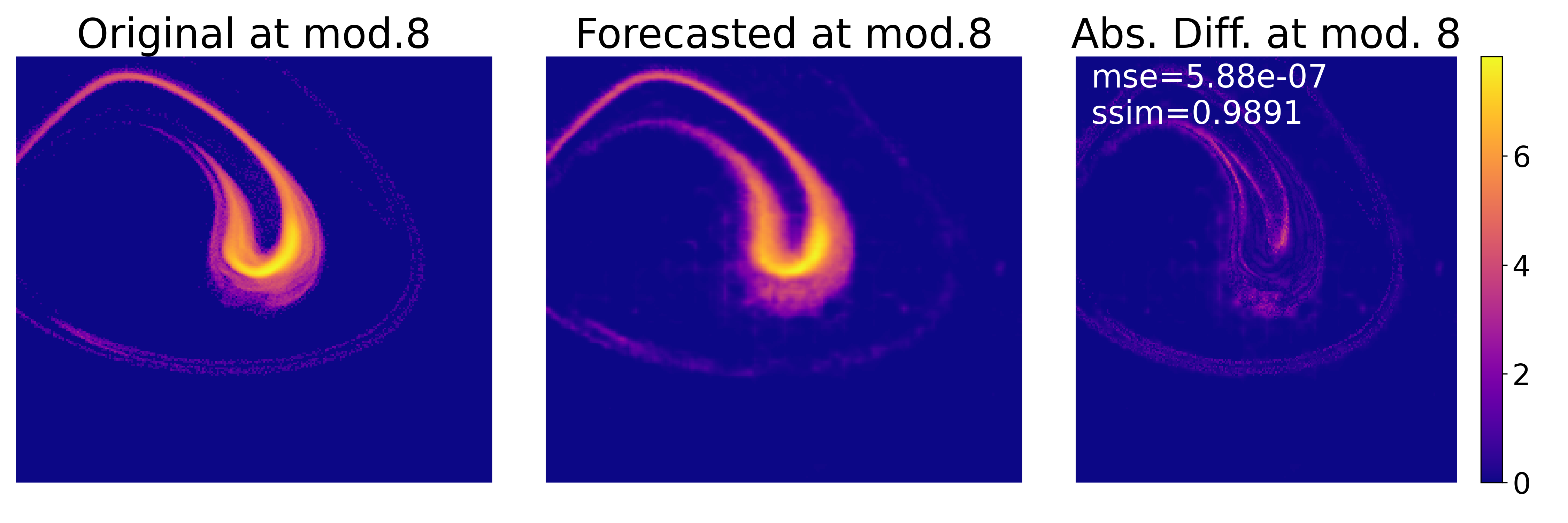}
    \end{minipage}
    \begin{minipage}[b]{0.48\linewidth}
        \centering
        \includegraphics[width=1.0\textwidth]{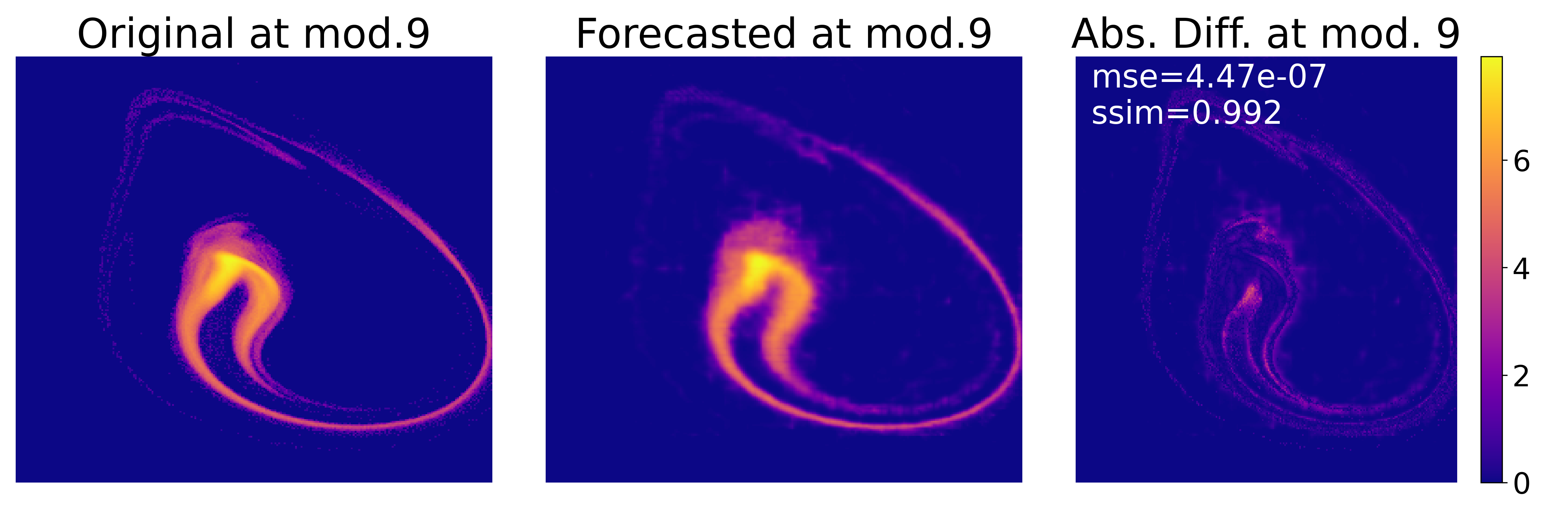}
    \end{minipage}
    \begin{minipage}[b]{0.48\linewidth}
        \centering
        \includegraphics[width=1.0\textwidth]{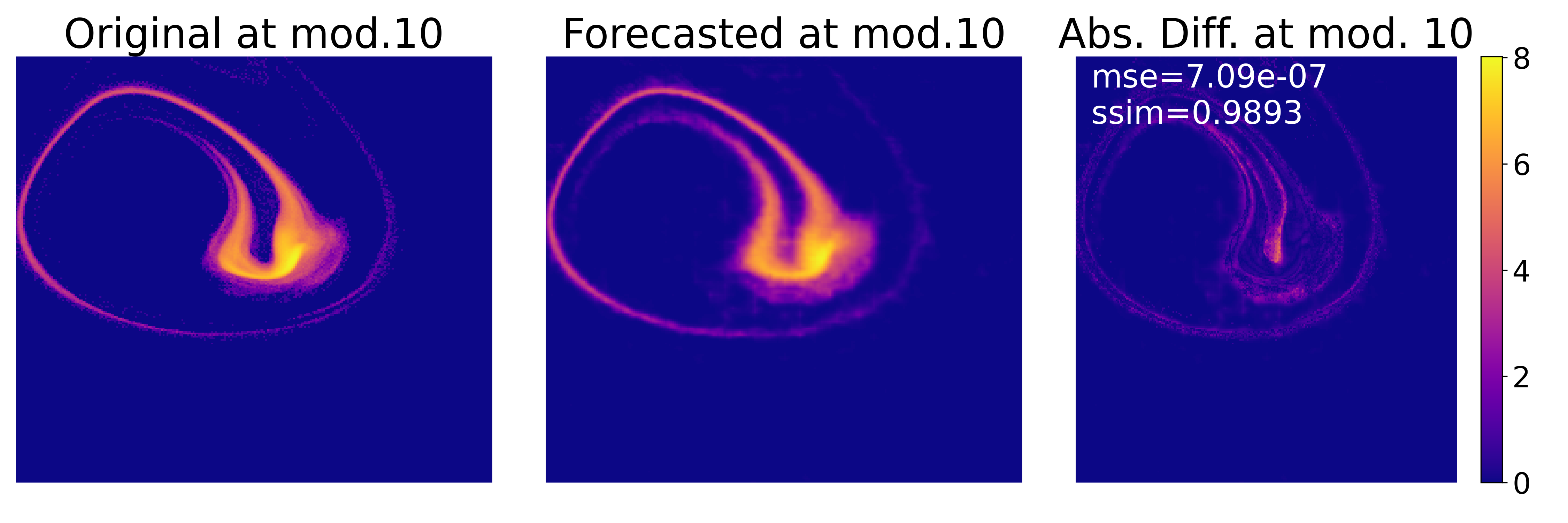}
    \end{minipage}
    \begin{minipage}[b]{0.48\linewidth}
        \centering
        \includegraphics[width=1.0\textwidth]{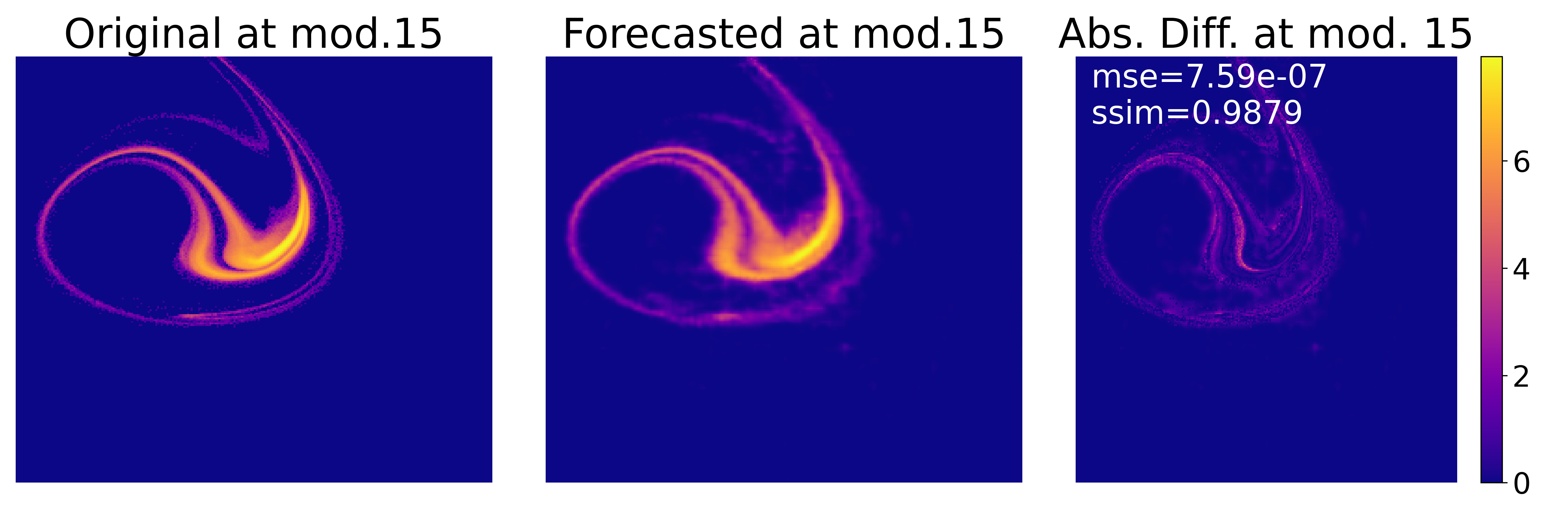}
    \end{minipage}
    \begin{minipage}[b]{0.48\linewidth}
        \centering
        \includegraphics[width=1.0\textwidth]{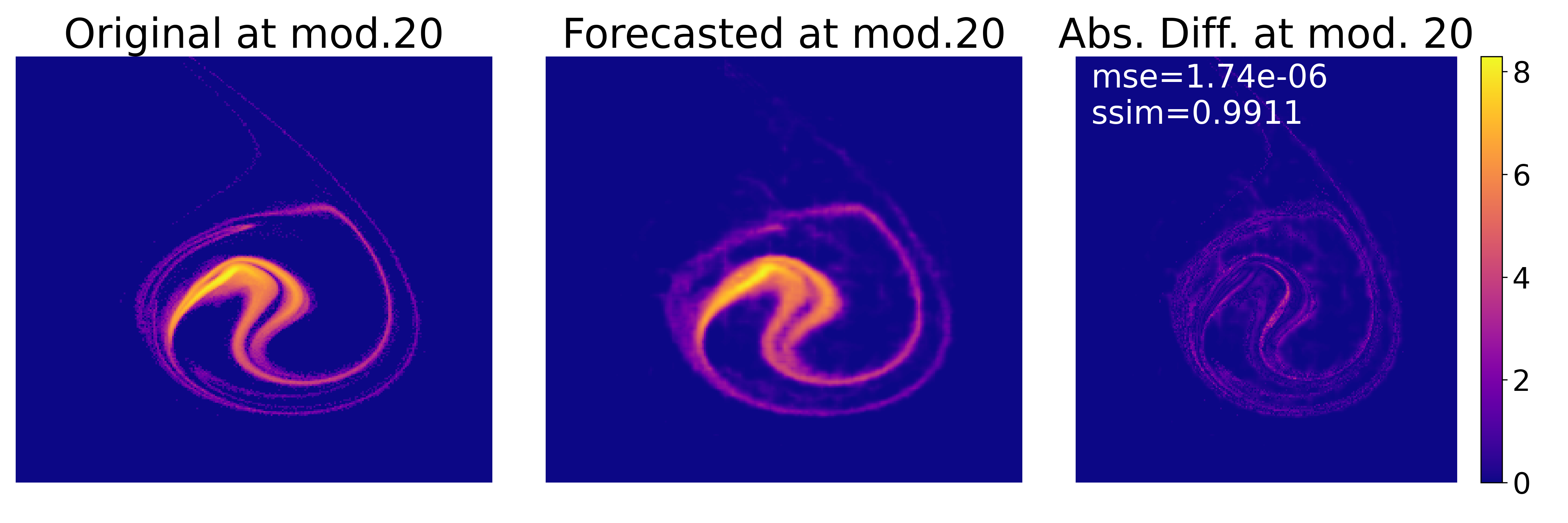}
    \end{minipage}
    \begin{minipage}[b]{0.48\linewidth}
        \centering
        \includegraphics[width=1.0\textwidth]{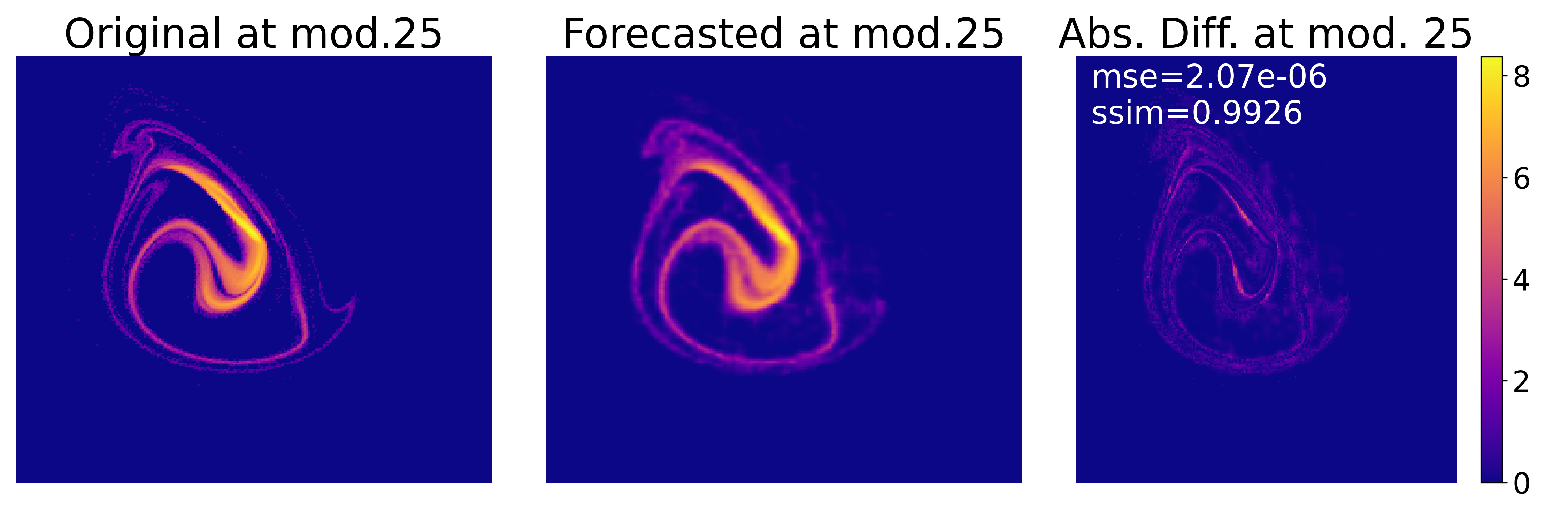}
    \end{minipage}
    \begin{minipage}[b]{0.48\linewidth}
        \centering
            \includegraphics[width=1.0\textwidth]{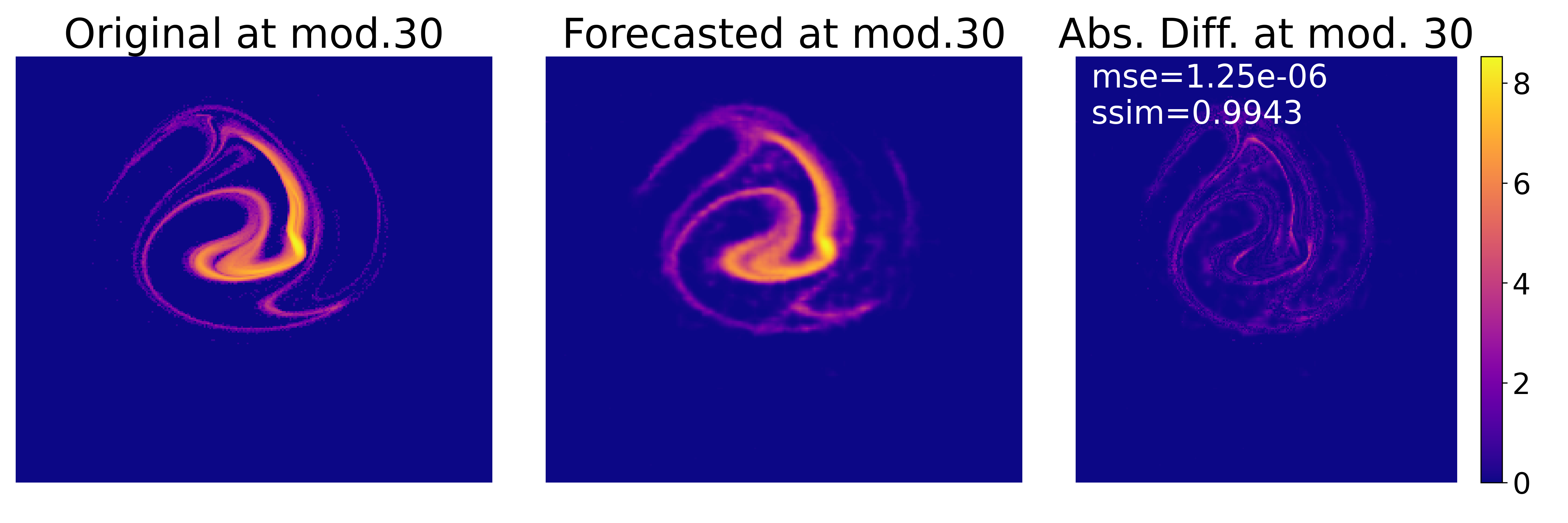}
    \end{minipage}
    \begin{minipage}[b]{0.48\linewidth}
        \centering
        \includegraphics[width=1.0\textwidth]{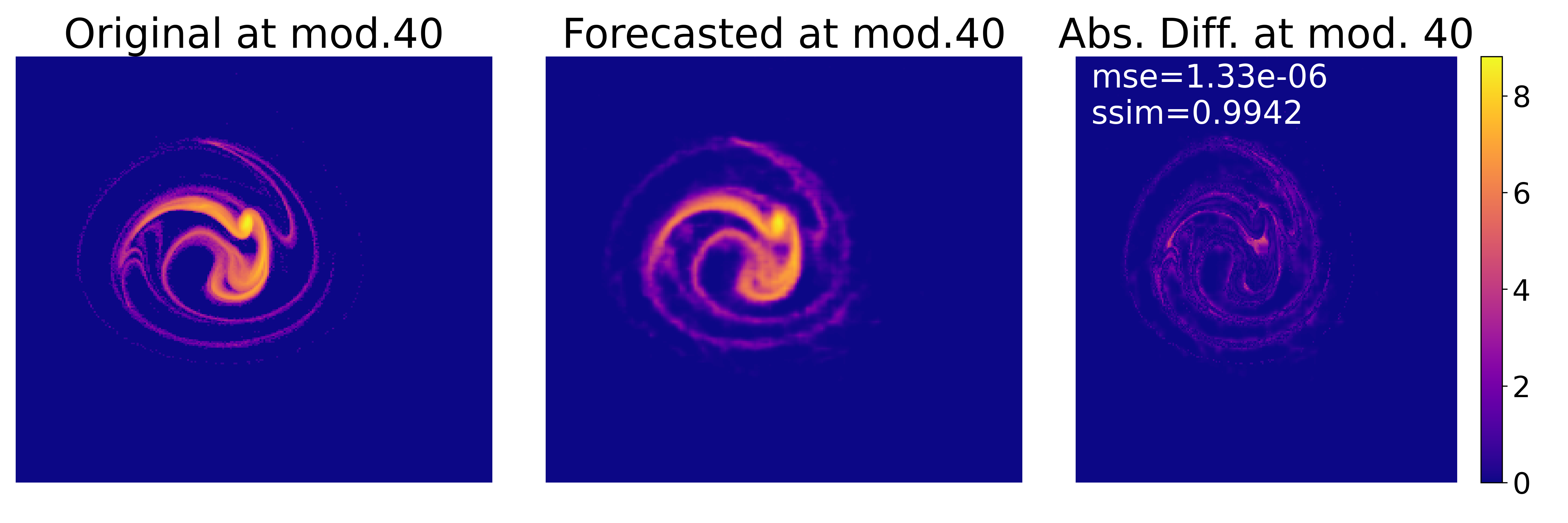}
    \end{minipage}
    \begin{minipage}[b]{0.48\linewidth}
        \centering
        \includegraphics[width=1.0\textwidth]{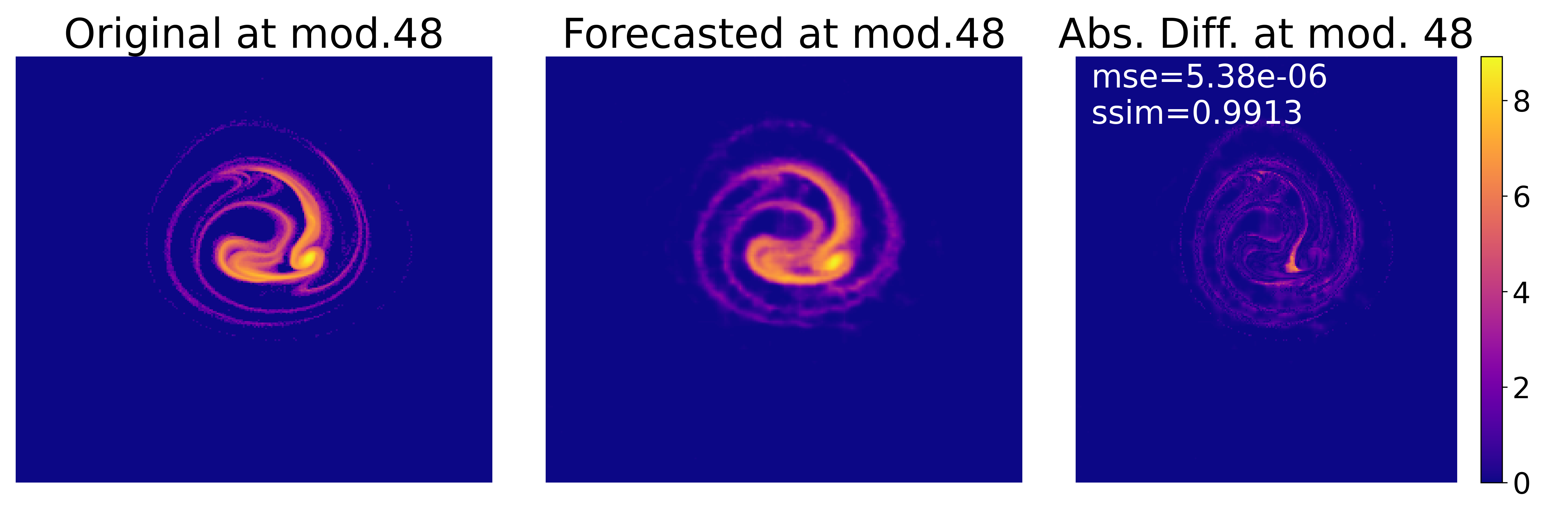}
    \end{minipage}
    \begin{minipage}[b]{1.0\linewidth}
        \centering
        \includegraphics[width=0.8\textwidth]{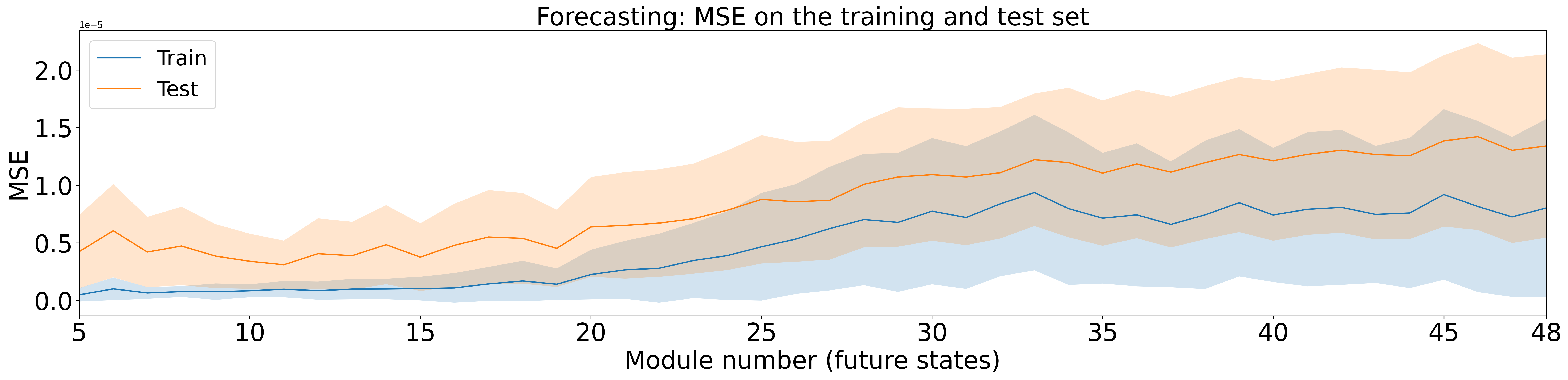}
    \end{minipage}
    \begin{minipage}[b]{1.0\linewidth}
        \centering
        \includegraphics[width=0.8\textwidth]{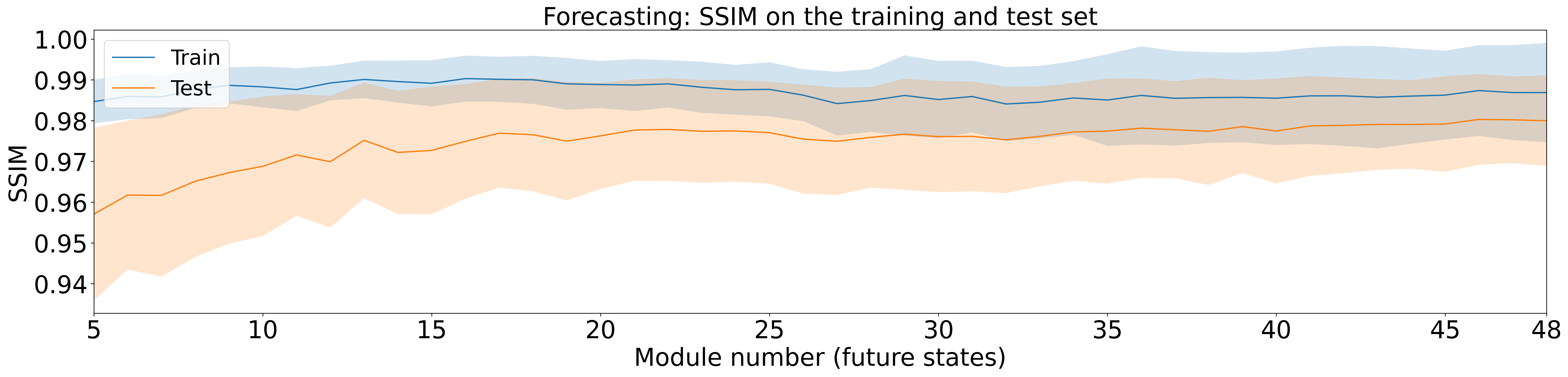}
    \end{minipage}
    \caption{Forecasting results: Forecasted projections (shown $E-\phi$ only) across different modules given first four projections as inputs. The original projection is presented against the forecasted along with the absolute difference between both. The MSE and SSIM for the entire training and test set are plotted as a shaded plot where the the central line is the mean and the boundaries of the region defines the standard deviation.}
    \label{fig:forecastingability}
\end{figure}

In the last step of the process, the forecast latent points are passed through the decoder of the CVAE to reconstruct the phase space projections. In Fig.~\ref{fig:forecastingability}, we showcase the $E-\phi$ projection (see supplementary Figs. 18 and 19 for x-y and x'-y' projections) using the initial four phase space projections from the test set as input. The figure depicts the original projection, the predicted projection, and their absolute difference at various modules. The corresponding MSE and SSIM for the projection reveal high similarity, with MSE of the order of $10^{-7}$ and SSIM exceeding 0.99. It is also seen that there is an increasing trend in MSE values across modules whereas SSIM does not vary much. Observing the absolute difference plots, we note minimal discrepancies, primarily around the tails of the forecasted projection for initial modules. However, discrepancies grow for later modules, evident in higher MSE values. This growth in MSE is expected as the LSTM's inputs are the true values of $M_1-M_4$, based on which it predicts an estimate $\hat{M_5}$ of $M_5$ and then uses its own prediction to generate $\hat{M_6}$ and so on in an iterative manner in which the errors introduced by the CVAE and LSTM are propagated and accumulated leading to a continuous increase in error.

This expected growth in error with forecasting distance is sumarized by calculating the average MSE and SSIM over the entire training and test sets (including all projections) for each of the modules, as shown at the bottom of Fig.~\ref{fig:forecastingability}. The central lines in the plots are the mean values, whereas the boundaries of the shaded region are the standard deviations. It can be clearly seen that due to the autoregressive nature of the model and error accumulation, the MSE builds up and gradually rises towards the later modules. The MSE and SSIM plots provide promising values confirming the good performance of CLARM on forecasting. The forecasting of all the projections in all the modules takes less than one second whereas HPSim takes around 10 minutes with similar computing infrastructure, resulting in a speed up by a factor of $\sim 600$. 

The exceptional computational speed up of the CLARM method makes the method extremely well suited for various real-time accelerator applications. The CLARM method can be used as a virtual diagnostic in which CLARM predicts a detailed evolution of the beam's phase space through the entire LANSCE accelerator based on the current RF module settings and using only 4 initial steps from the much slower HPSim physics-based model as its initial points. This fast virtual diagnostic gives the accelerator operators a virtual view of the beam's behavior so that they can understand which accelerator parameters must be adjusted in order to correct issues like beam spill or to shape the beam's phase space for specific experiments. Furthermore, because of its speed, the virtual CLARM-based diagnostic can quickly be iterated in order to find an optimal accelerator tune of the first 4 Module settings virtually before implementing those settings on the real machine. In general, the application of such an approach to any large accelerator will provide a substantial benefit for simulating beam dynamics and for accelerator optimization.

In this section, we have shown reconstruction, visualization, generation and forecasting results. One detail to note is that the dataset contains phase space projections at the beginning of every module. The length of the accelerating section (or module) is usually around a few meters, depending on the accelerator. So, currently, we are observing the phase space of charged particles once every 3-10 meters. If we want to collect data at different positions within a single module, the particle tracking simulations can take an enormous amount of time. The advantage of using generative models is that they can be used for interpolating the phase space projections between two modules \cite{berthelot2018understanding}. It can convert low-resolution simulations (along z) to high-resolutions, which becomes demanding while simulating charged particle dynamics in larger accelerators. We have not studied interpolation aspects in this research, but it is a goal of future research. We have also mentioned that uncertainty quantification is a byproduct of probabilistic models like VAE. By sampling the latent space for the first few modules, the LSTM and decoder can be used to generate phase space projections in all the modules. A detailed investigation of the uncertainty analysis aspect is a part of future research work once the model is expanded to predict over a much larger data set in which the settings of all 48 modules are sampled.

\section{Conclusions}\label{sec: conclusions}
Overall, our paper presents a novel approach for learning spatiotemporal dynamics and its application is shown for the complex dynamics of charged particle beams in linear accelerators. We develop a conditional latent autoregressive recurrent model for the generation and forecasting of phase space projections. The proposed model consists of CVAE and LSTM to learn spatial and temporal dynamics independently but combines them together in an autoregressive framework. The model is capable of generating realistic projections at various accelerator locations by sampling the latent space and decoding. The model can forecast projections in later modules given initial modules without any supervision from RF field set points. The results are promising when tested on a variety of evaluation metrics for reconstruction, generation, and forecasting ability. We have performed a detailed visualization of the latent space through higher-dimensional plots and its projections, PCA, t-SNE, and UMAP. The proposed methodology brings computational speed, robustness, and enhanced interpretability for solving spatiotemporal dynamics problems. This general method provides a computational speed up of approximately 600x for complex beam dynamics and is applicable to a wide range of accelerator tuning, optimization, and virtual diagnostics applications. 

\section{Availability of data and materials}
The data used in this article is available on Zenodo at \url{https://zenodo.org/10.5281/zenodo.10819001}.

\section{Code availability}
The code used in this article is available on GitHub at \url{https://github.com/mahindrautela/CLARM}.

\nocite{}
\bibliography{sn-bibliography}

\section{Acknowledgements}
Los Alamos National Laboratory (LANL) is operated by Triad National Security, LLC, for the national Nuclear Security Administration of the US Department of Energy (Contract No. 89233218CNA000001). This work was supported by the LANL LDRD Program Directed Research (DR) project 20220074DR.

\section{Authors' contributions}
Conceptualization (M.R., A.S.), Methodology (M.R., A.S.), Data curation (M.R., A.W.), Formal analysis (M.R., A.S.), Investigation (M.R., A.W, A.S.),  Writing – original draft (M.R.), Writing – review and editing (M.R., A.W, A.S.), Funding acquisition (A.S.), Supervision (A.S.).

\section{Additional information}
\subsection{Supplementary information}
The supplementary information is available with this article.
\section{Competing Interests}
The authors declare no competing interests

\end{document}